\documentclass[aps,twocolumn,floats,nofootinbib]{revtex4}
\usepackage{graphics,graphicx,epsfig}
\usepackage{amssymb,color}
\usepackage{epsf,epstopdf}
\usepackage{amsmath}
\usepackage{pdfpages}
\usepackage{natbib}
\usepackage{bibentry}

\usepackage[paperwidth=8.5in,paperheight=11in,centering,hmargin=0.6in,vmargin=0.6in]{geometry}

\newcommand{\avg}[1]{\left< #1 \right>} 

\begin{document}

\title{Diverse spatial expression patterns emerge from \\ common transcription bursting kinetics}
\author{Benjamin Zoller$^a$\footnote{These authors contributed equally.}, Shawn C. Little$^{b *}$, Thomas Gregor$^a$\footnote{Correspondence: tg2@princeton.edu}}
\affiliation{$^a$Joseph Henry Laboratories of Physics and the Lewis Sigler Institute for Integrative Genomics, Princeton University, Princeton, NJ 08544, USA \\ $^b$Department of Cell and Developmental Biology, University of Pennsylvania Perelman School of Medicine, Philadelphia, PA 19143, USA}
\date{\today}

\begin{abstract}
In early development, regulation of transcription results in precisely positioned and highly reproducible expression patterns that specify cellular identities. How transcription, a fundamentally noisy molecular process, is regulated to achieve reliable embryonic patterning remains unclear. In particular, it is unknown how gene-specific regulation mechanisms affect kinetic rates of transcription, and whether there are common, global features that govern these rates across a genetic network. Here, we measure nascent transcriptional activity in the gap gene network of early \emph{Drosophila} embryos and characterize the variability in absolute activity levels across expression boundaries. We demonstrate that boundary formation follows a common transcriptional principle: a single control parameter determines the distribution of transcriptional activity, regardless of gene identity, boundary position, or enhancer-promoter architecture. By employing a minimalist model of transcription, we infer kinetic rates of transcriptional bursting for these patterning genes; we find that the key regulatory parameter is the fraction of time a gene is in an actively transcribing state, while the rate of Pol II loading appears globally conserved. These results point to a universal simplicity underlying the apparently complex transcriptional processes responsible for early embryonic patterning and indicate a path to general rules in transcriptional regulation.
\end{abstract}

\maketitle



\section{Introduction}

A central question in gene regulation concerns how discrete molecular interactions generate the continuum of expression levels observed at the transcriptome-wide scale \citep{Lionnet:2012hi}. A large set of molecular activities are required to elicit RNA transcription, including transcription factor DNA binding, chromatin modifications, and long-range enhancer-promoter interactions \citep{Voss:2013ca,Levine:2014bn}. However, in most cases it is unclear which, if any, of these interactions predominantly regulate the RNA synthesis rate and the associated variability for any given gene \citep{Stavreva:2012ib}. In general, for genes whose transcription rates depend on levels of external inputs, we do not know which regulatory steps are preferably tuned to achieve functional levels of mRNA expression. Overall, it is unknown whether constraints exist that might select common mechanisms for modulating transcriptional activity across genes, space and time. 

Addressing these questions requires measuring the kinetic rates of transcription, in absolute units. Several studies using single molecule counting approaches have documented the inherently stochastic nature of transcription \citep{Raj:2006gq,Zenklusen:2008co,Taniguchi:2010cb,Little:2013dr}. In organisms ranging from bacteria to vertebrates, genes exhibit transcriptional bursts characterized by intermittent intervals of mRNA production followed by protracted quiescent periods \citep{Golding:2005dm,Suter:2011hk,Bothma:2014jq}. This inherent stochasticity in gene activation results in greater cell-to-cell variability than expected from models of constitutive expression (Poisson noise) \citep{Blake:2003kp}. Transcriptional bursting and the variability associated with it is well-described and formalized by a simple telegraph or two-state model of transcription in which a locus alternates at random between active and inactive states \citep{Peccoud:1995ww}. Despite the prevalence of transcriptional bursting, in the majority of cases it is unknown what processes determine its kinetics \citep{Hebenstreit:2013hy,Tantale:2016fb}. Nor is it widely understood how kinetic rates are controlled by external input signals (Molina et al, 2013; Senecal et al, 2014). By means of quantitative modeling and properly calibrated experiments, it is possible to acquire an understanding of the mechanisms underlying transcription regulation based on their signature in the measured transcriptional variability or noise \citep{Tkacik:2008ht,Sanchez:2011ic,Munsky:2012ie,Jones:2014cl,Rieckh:2014ic,Zoller:2015im}.

\emph{Drosophila} embryos provide an ideal model to explore the relationship between intrinsic regulatory inputs and transcriptional output \citep{Gregor:2014ev}. Early embryos express many genes in graded patterns of transcription levels in response to modulatory inputs \citep{Struhl1992}. Spatial domains, where gene expression levels transition from highly active to nearly silent, are functionally the most critical for the developing embryo, as they eventually determine specification of cell identities \citep{Gergen1986,Kornberg1993}. Among the earliest expressed genes in \emph{Drosophila} development are the gap genes, which encode transcription factors responsible for anterior-posterior (AP) patterning \citep{Jaeger:2011gp}. Each gap gene is expressed in its own unique pattern, and their expression boundaries arise at distinct and precise positions along the AP axis \citep{Dubuis:2013cw}. Gene expression levels near these boundaries are spatially graded across several cell diameters, and the intermediate levels of these gap genes confer patterning information necessary for embryonic segmentation \citep{Gergen1986,Kornberg1993,Yu2008,Dubuis:2013cp}. Thus the precise control of expression levels across these boundaries is essential for properly patterned cell fate specification. 

The regulation of the gap genes appears highly complex. Many activating and repressing factors determine expression boundaries through complex layers of homo- and heterotypic protein interactions at multiple promoters and enhancers \citep{Driever1989,Kraut:1991wk,Simpson-Brose1994,Sauer1995,Lebrecht2005,Perry:2011cb}. Given this complexity, an intuitive expectation is that expression rates emerge from carefully tuned transcription factor concentrations and binding affinities. The expression pattern of each gap gene results from the interplay between multiple enhancers responsible for the spatio-temporal regulation of transcription \citep{Perry:2011cb,Kvon:2014fn}. Each enhancer possesses distinct arrangements of binding sites that enable cooperation, as well as competition between multiple activators and repressors \citep{Segal:2008fi}. The collective activity of cross-regulating transcriptional modulators generates rates of gene expression that vary with position in the embryo, thereby forming expression boundaries \citep{Jaeger:2004ko,Manu:2009fl,Briscoe:2015js}. Thus a straightforward prediction is that the underlying bursting kinetics will differ between boundaries. However, no previous work has investigated how kinetic rates compare between genes or across expression boundaries.

To address these questions, we developed a single molecule fluorescent in situ hybridization (smFISH) based approach that allows highly accurate counting of nascent RNAs molecules in individual nuclei. Measurements of the major \emph{Drosophila} gap genes provide access to absolute mean transcription levels and their variability across loci. Using a simple telegraph model to interpret these measurements provides insights into the underlying transcription processes and reveals a common basic principle that unifies the transcriptional output of all genes at all positions in the embryo. Finally, based on this model we estimate its kinetic parameters from the data and determine which and how these change as a function of position. Surprisingly, despite the diversity of cis-regulatory architecture and trans-acting factors regulating these genes, we established that transcriptional bursting is tightly constrained across all expression boundaries. Our findings suggest an underlying emerging simplicity in the transcriptional regulation of the apparently complex process of embryo segmentation.

\section{Results}

\subsection*{Precise measurements of transcriptional activity}

\begin{figure*}
\centerline{\includegraphics[scale=1.0]{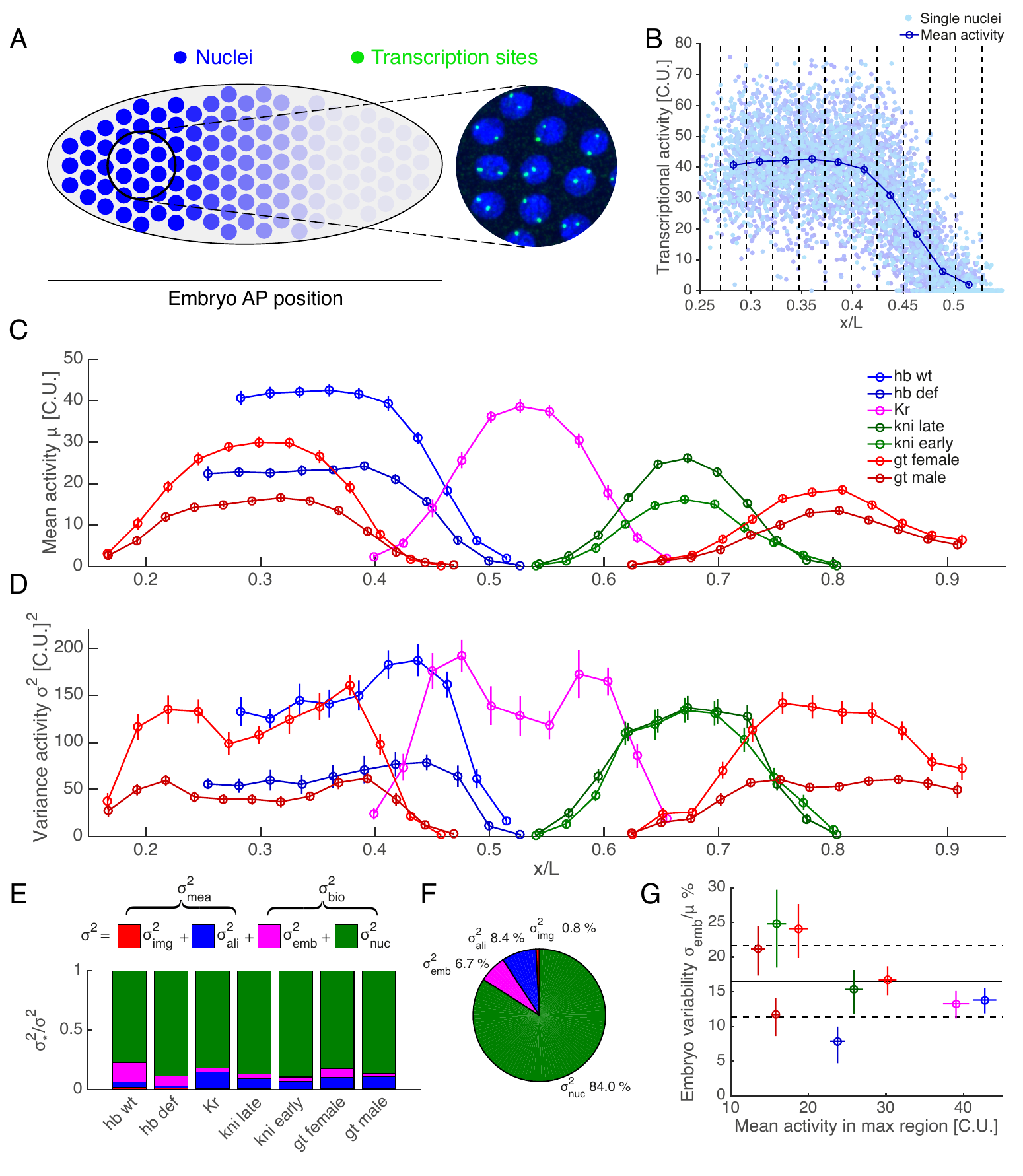}}
\caption{{\bf Absolute quantification of gap gene transcriptional activity.}
A) Transcriptional activity of individual nuclei measured by single molecule mRNA-FISH in nuclear cycle 13 of the blastoderm embryo. The activity of individual nuclei results from the summed intensity of each locus (transcription sites, green bright spots) normalized by the average intensity of a single fully elongated cytoplasmic mRNA (cytoplasmic unit, C.U.). 
B) Transcriptional activity profile for the gap gene \emph{hb} as a function of AP position in \% egg length for 18 embryos. Each dot represents the total mean intensity of nascent transcription activity in C.U. for individual nuclei. Vertical dashed lines define AP bins covering 2.5\% of egg length; crosses display mean activity in each bin .
C) Mean transcriptional activity as a function of AP position for all measured trunk gap genes in C.U.. 
D) Total variance of transcriptional activity as a function of AP position for all measured trunk gap genes in [C.U.]$^2$.
E) Decomposition of the total variance $\sigma^2$ into measurement error and biological variability for all genes. Estimates of imaging error (red) alignment error (blue), embryo-to-embryo variability (magenta) are decoupled from the total variance. The remaining variance corresponds to biological variability and is termed intrinsic nucleus-to-nucleus variability in the text (green).
F) Global decomposition of total variance for the entire data set. Nucleus-to-nucleus variability largely dominates in the blastoderm embryo.
G) Fractional embryo-to-embryo variability (CV) as a function of mean activity (solid black line: mean ratio; dashed lines: 66\% confidence intervals) reaches 16$\pm$4\% (CV) in the maximally expressed regions that are the most reproducible across embryos. Notably, this corresponds to absolute reproducibility as each embryo peaks at comparable mean values.
Error bars stand for the 66\% confidence intervals.}
\label{fig:01}
\end{figure*}

During early fly development, the formation of gene expression boundaries arises from spatially varying transcription factor concentrations. Early embryos thus provide a natural, experimentally accessible context in which to ask how input factors shape transcription dynamics. Here we performed single molecule fluorescent in situ hybridization (smFISH) \citep{Little:2013dr}, namely we used fluorescent oligonucleotide probes to label single mRNA molecules in fixed embryos and estimate the intensity of transcription sites (co-localized nascent transcripts) and individual cytoplasmic mRNAs. This method enables us to measure instantaneous transcriptional activity of gene loci per nucleus in terms of ``cytoplasmic unit'' intensity (C.U.), by normalizing the total fluorescent intensity of each locus to the equivalent number of processed cytoplasmic mRNA molecules (Fig. 1A, B). Our enhanced method increases sensitivity by 3- to 4-fold (see Methods), thereby enabling precise counting of nascent transcripts, and hence measurement of transcriptional activity across expression boundaries.

We measured the transcriptional activity per nucleus for all the four trunk gap genes \emph{hunchback} (\emph{hb}), \emph{Kr\"uppel} (\emph{Kr}), \emph{knirps} (\emph{kni}), and \emph{giant} (\emph{gt}) along the embryo AP axis. These genes are expressed early during development in broad spatial domains, thus permitting transcription measurements from thousands of synchronized nuclei across relatively small numbers of embryos; all factors that favor low measurement error (Fig. 1C and 1D, N$>$10 embryos per combination of genes and genotype). Expression level analysis during the mid- to late portion of interphase 13 ensures that sufficient time has elapsed to allow these genes to attain steady-state levels of transcribing RNA polymerase II (Pol II; Supplement, Fig. S1); and DNA replication occurs in early interphase for these loci \citep{Yuan:2014da}, such that observations during later times eliminate ambiguity arising from variable numbers of transcriptionally active loci. Since loci on recently duplicated chromatids are often closely apposed in space, we measure total transcription per nucleus (as previously described in Little et al, 2013) and use these data to infer the statistical properties of individual loci. As a control, we generated data from embryos heterozygous for a \emph{hb} deficiency, and observed half the wild-type level of expression per nucleus (Fig. 1C). Importantly, we also observe a corresponding decrease in variance to half that observed in wild-type (Fig. 1D), supporting previous findings that all loci behave independently \citep{Little:2013dr}. These results demonstrate the suitability of using total transcriptional activity per nucleus to infer the behavior of individual loci. 
 
Since biological variance greatly constraints models of the regulatory processes underlying transcription, we need to determine how the total observed gene expression variability decomposes into measurement error, embryo-to-embryo differences, and intrinsic fluctuations in individual nuclei. We assessed the performance of our measurements with a two-color smFISH approach, labeling each mRNA in alternating colors along the length of the mRNA strand. This approach allowed us to perform an independent normalization in each channel, and thus to characterize the various sources of measurement error, such as noise stemming from imaging, normalization, and embryo alignment along the anterior-posterior axis (Supplement, Fig. S2). For all genes and at all positions, the measurement variability represents less than 10\% of the total variance on average (Fig. 1E), indicating that biological variability is largely dominant in our measurements \citep{Dubuis:2013cw}. Importantly, this variability arises almost entirely from differences between nuclei, rather than differences between embryos (Fig. 1F). Notably, the low embryo-to-embryo variability in the maximally expressed regions (16$\pm$4\% CV, Fig. 1G) emphasizes that the mean expression levels across embryos are reproducible in absolute units (Fig. 1C). Thus the measured expression noise mainly stems from zygotic transcription, and is intrinsic to the molecular processes underlying transcription rather than from an extrinsic source of variability, such as maternal age or live history, or other processes occurring during oogenesis. Our low measurement error combined with largely dominant intrinsic variability facilitate the analysis of the noise-mean relationship and permit the inference of expression kinetics from several hundred nuclei at each position along the AP axis (Fig. 1B), as detailed below.

\subsection*{Single parameter distribution of transcriptional activity across all expression boundaries}

The expression patterns of the four gap genes are each determined by multiple enhancer elements positioned at varying distances from their promoters \citep{Perry:2011cb,Kvon:2014fn}. In addition, each enhancer contains a variable number of binding motifs for multiple patterning input factors with cross-regulatory interactions \citep{Schroeder:2004cs,OchoaEspinosa:2005ej}. These features, as well as evidence from genetic manipulations \citep{Hoch:1990va,JACOB:1991tq,Pankratz:1992up}, indicate that many molecular processes are required to regulate the transcription rates that generate observed mRNA expression levels with their stereotypical modulation as a function of position in the embryo (Fig. 1C). Given the diversity of input factors and molecular control elements involved in the transcription process, it would appear likely that different genes exhibit vastly different and uniquely defined transcription kinetics. In order to make progress in our understanding of these complex multi-factorial relationships, we capitalize on the fact that the kinetics of the processes underlying transcription determine not only these mean expression levels but also the observed variability in our data (Fig. 1D). Thus we can use measurements of the noise-mean relationship to characterize the expression kinetics for individual genes. 

To characterize the different noise-mean relationships in our system, we examined the dependence of variability on mean transcription levels (Fig. 2A). In agreement with previous measurements \citep{Little:2013dr}, genes span a similar dynamic range of expression levels across boundaries, spanning nearly zero to a maximum value of 34$\pm$6 C.U. across genes (Fig. 1C). In addition, transcription is inherently variable: at all positions and across all genes, variability exceeds that expected from a simple model of constitutive activity (Poisson noise), with noise (measured as CV$^2$) approximately 12 times larger than Poisson noise for a mean transcriptional activity below 10 C.U. (Fig. 2A). However, the noise-mean relationship follows an unexpectedly similar overall trend for all genes (Fig. 2A). First, unlike many other systems (bacteria, yeast, mammalian culture cells), there is no clearly identifiable noise floor at high expression \citep{Taniguchi:2010cb,Keren2015,Zoller:2015im}. The absence of such an extrinsic noise floor is likely a key feature of early embryo development since nuclei are highly synchronized (cell-cycle wise) and share the same environment (syncytial blastoderm). Second, the apparent collapse of each gene on a unique curve is unexpected and atypical given the different promoter-enhancer architectures \citep{Hornung:2012jg,Sanchez:2013fp}. Even more strikingly, when we converted the units from activity in C.U. to Pol II number counts g (i.e. by taking into account individual gene length, copy number, and probe configuration, see Supplement), the 2$^{\rm nd}$, 3$^{\rm rd}$ and 4$^{\rm th}$ cumulants for all genes are almost uniquely determined by a single parameter, the mean activity (Fig. 2B-D). Thus transcriptional activity is characterized across the entire expression range and for all genes by a unique common single-parameter distribution exclusively specified by the mean. Such a universal feature suggests that despite the well-documented diversity of cis-regulatory elements and trans-acting factors, a common conserved set of processes is regulated to determine transcription kinetics across nearly all expression boundaries in the early embryo.

\begin{figure*}
\centerline{\includegraphics[width = \linewidth]{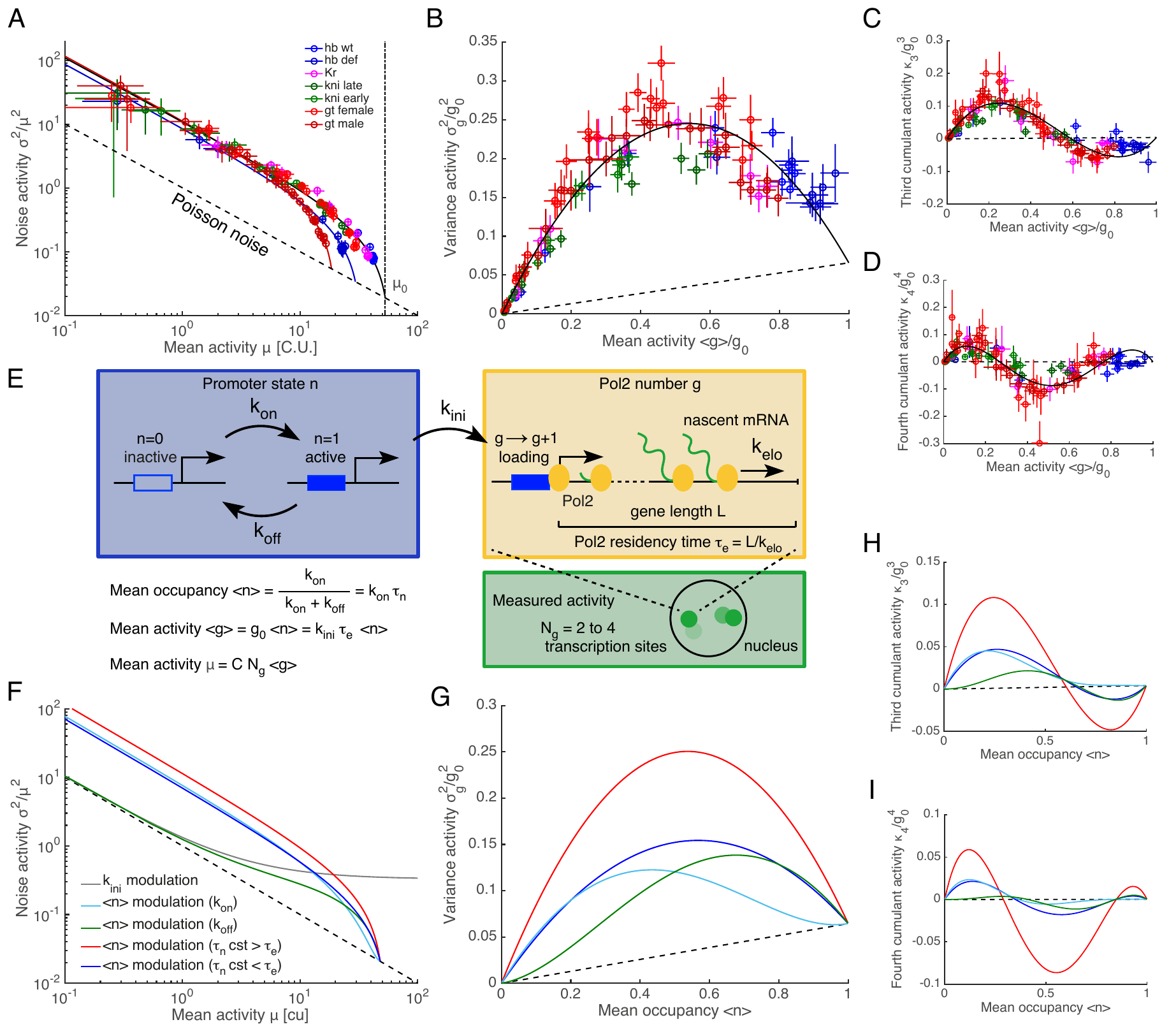}}
\caption{{\bf A two-state model recapitulates single-parameter modulation for all genes.}
A) Noise-mean relationship for gap genes, with noise defined as the fraction between variance and the squared mean activity (CV$^2$). Dashed line stands for the Poisson background, corresponding to the lowest attainable noise. Solid lines were obtained by fitting $2^{\rm nd}$ order polynomials for each gene. The collapse of the trend to Poisson noise at high expression suggests an upper limit $\mu_0$ in attainable expression levels (vertical dashed line). 
B-D) Normalized $2^{\rm nd}$, $3^{\rm rd}$ and $4^{\rm th}$ cumulant as a function of normalized Pol II counts for a single gene copy. The activity intensities in C.U. were converted into Pol II counts g by taking into account actual fluorescent probe locations and gene lengths. Assuming independence, the mean and the cumulant were divided by the gene copy number $N_g=2,4$. Dashed lines stand for the Poisson background. The solid lines were obtained by fitting the cumulants with $2^{\rm nd}$, $3^{\rm rd}$ and $4^{\rm th}$ order polynomial, constrained to match the Poisson level at maximum Pol II counts g0.
E) Two-state model for statistical properties of measured transcriptional activity: promoter switches stochastically between an inactive and active state leading to intermittent Pol II initiation events. The mean activity in Pol II counts is $\avg{g} = k_{\rm ini} \tau_e \avg{n}$, with initiation rate $k_{\rm ini}$, elongation time $\tau_e=L/k_{\rm elo}$, and promoter mean occupancy $\avg{n} = k_{\rm on}/(k_{\rm on} +k_{\rm off}) \in [0,1]$. The maximal Pol II count is given by $g_0 = k_{\rm ini} \tau_e$. The measured mean activity in C.U. is $\mu = C N_g \avg{g}$, where $N_g$ is the gene copy number and $C \in [0,1]$ a conversion factor that depends on the probe locations on transcripts.
F) Noise-mean relationship and G-I) normalized $2^{\rm nd}$, $3^{\rm rd}$ and $4^{\rm th}$ cumulants predicted by the two-state model under different single parameter mean activity modulation schemes: Pol II initiation rate $k_{\rm ini}$ (gray), OFF-rate $k_{\rm off}$ (green), ON-rate $k_{\rm on}$ (cyan), and promoter occupancy $\avg{n}$ at constant switching correlation time $\tau_n = (k_{\rm on}+ k_{\rm off})^{-1}$ (red for $\tau_n>\tau_e$, and blue for $\tau_n<\tau_e$). Notably, $\avg{n}$ modulation at slow switching ($\tau_n>\tau_e$) achieves numerical values that closely match the trends of our data.
Error bars stand for the 66\% confidence intervals.}
\label{fig:02}
\end{figure*}

\subsection*{Two-state model identifies unique control parameter underlying universality}

These shared features suggest that a common general model can describe the regulated kinetics of all genes. A popular minimalist model that accounts for intrinsic super-Poissonian variability is the two-state model, where loci switch stochastically between transcriptionally inactive and active states, with transcription initiation only occurring in the latter (Fig. 2E) \citep{Peccoud:1995ww}. The two-state model has been widely used to describe the distribution of mature mRNA and protein counts \citep{BarEven:2006dz,Raj:2006gq,Zenklusen:2008co}. Such a simple mechanistic model enables estimation of switching rates between promoter states ($k_{\rm on}$ and $k_{\rm off}$) as well as the effective initiation rate $k_{\rm ini}$ \citep{Suter:2011hk,Larson:2013jt,Senecal:2014dz}. Our measurements of nascent transcript counts represent instantaneous transcriptional activity of Pol II molecules engaged in transcription, and thus provide a more direct measurement of instantaneous transcription activity compared to counts of mature mRNAs or proteins. For these reasons the two-state model presents a straightforward and parameter-sparse means to describe how discrete randomly occurring events generate a continuum of expression rates. It allows us to predict the dependence of variability on mean activity for different scenarios of parameter modulation, and to determine which of the kinetic rate parameters ($k_{\rm on}$, $k_{\rm off}$ and $k_{\rm ini}$) is modulated to form gene expression boundaries.

Given that the first four cumulants of our data are almost uniquely determined by a single parameter, we sought to only explore single parameter modulation consistent with the data. In addition, we assumed the Pol II elongation rate, $k_{\rm elo}$, to be constant and identical for all genes \citep{Garcia:2013ha,Fukaya2017}. When we solve the master equation for such a model (see Supplement), a comparison of the predicted noise activity (Fig. 2F) with our data (Fig. 2A) under each scenario unequivocally eliminates a modulation of $k_{\rm ini}$. Indeed, solely varying $k_{\rm ini}$ would lead to saturation of noise at high activity, which is not observed. Instead, our measurements are consistent with modulation of the fractional mean promoter occupancy $\avg{n}$, defined as $\avg{n}=k_{\rm on}/(k_{\rm on}+k_{\rm off})$. (Note that here occupancy refers to the active or ``ON'' state and $\avg{n}$ is thus bound between zero and one.) This value represents the fraction of time spent in the active state and is equivalent to the probability of finding a locus in the active state. Modulation of the mean production rate is thus determined by $\avg{n}$ rather than the rate at which Pol II molecules enter into productive elongation. Tuning only the mean occupancy $\avg{n}$ thus uniquely describes the formation of all expression boundaries regardless of their position in the embryo.

In principle, within the two-state model, either or both of the rates $k_{\rm on}$ and $k_{\rm off}$ may be tuned to modulate the mean occupancy. Modulation of $k_{\rm off}$ alone can be ruled out, since such a scenario would not capture the noise properly below 10 C.U. (see Fig. 2F). To distinguish between the other scenarios, we calculated the higher cumulants predicted by the two-state model (Fig. 2F-I). Although modulation of $k{\rm on}$ alone may explain the noise and variance (Fig. 2F-G), it does not capture the $3^{\rm rd}$ and $4^{\rm th}$ cumulants well (Fig. 2H-I).  Alternatively, the model can be parameterized by $\avg{n}$ and the correlation time $\tau_n = (k_{\rm on}+ k_{\rm off})^{-1}$ i.e., the characteristic time-scale for changes in promoter activity. Thus, both switching rates $k_{\rm on}$ and $k_{\rm off}$ are fully determined by $\avg{n}$ and $\tau_n$:
\begin{equation*}
k_{\rm on}=\frac{\avg{n}}{\tau_n} \quad \text{and} \quad k_{\rm off}=\frac{1-\avg{n}}{\tau_n}
\end{equation*}
For this new set of parameters, we observed a good match to the data when $\tau_n$ is maintained fixed and $\avg{n}$ varies alone (Fig. 2G-I), consistent with a single parameter modulation. In addition, the modulation of $\avg{n}$ in a slow switching regime ($\tau_n \geq \tau_e$ the elongation time) captures the cumulants more closely than in a fast regime ($\tau_n < \tau_e$). Thus the two-state model predicts that in this system simultaneously increasing active and decreasing inactive periods controls gene activity. Modulation of $\avg{n}$ at constant $\tau_n$ for all boundaries and at all positions represents a common, universal regulatory mechanism independent of gene identity. 

\subsection*{Kinetic transcription parameters in absolute units}

\begin{figure*}
\centerline{\includegraphics[width = \linewidth]{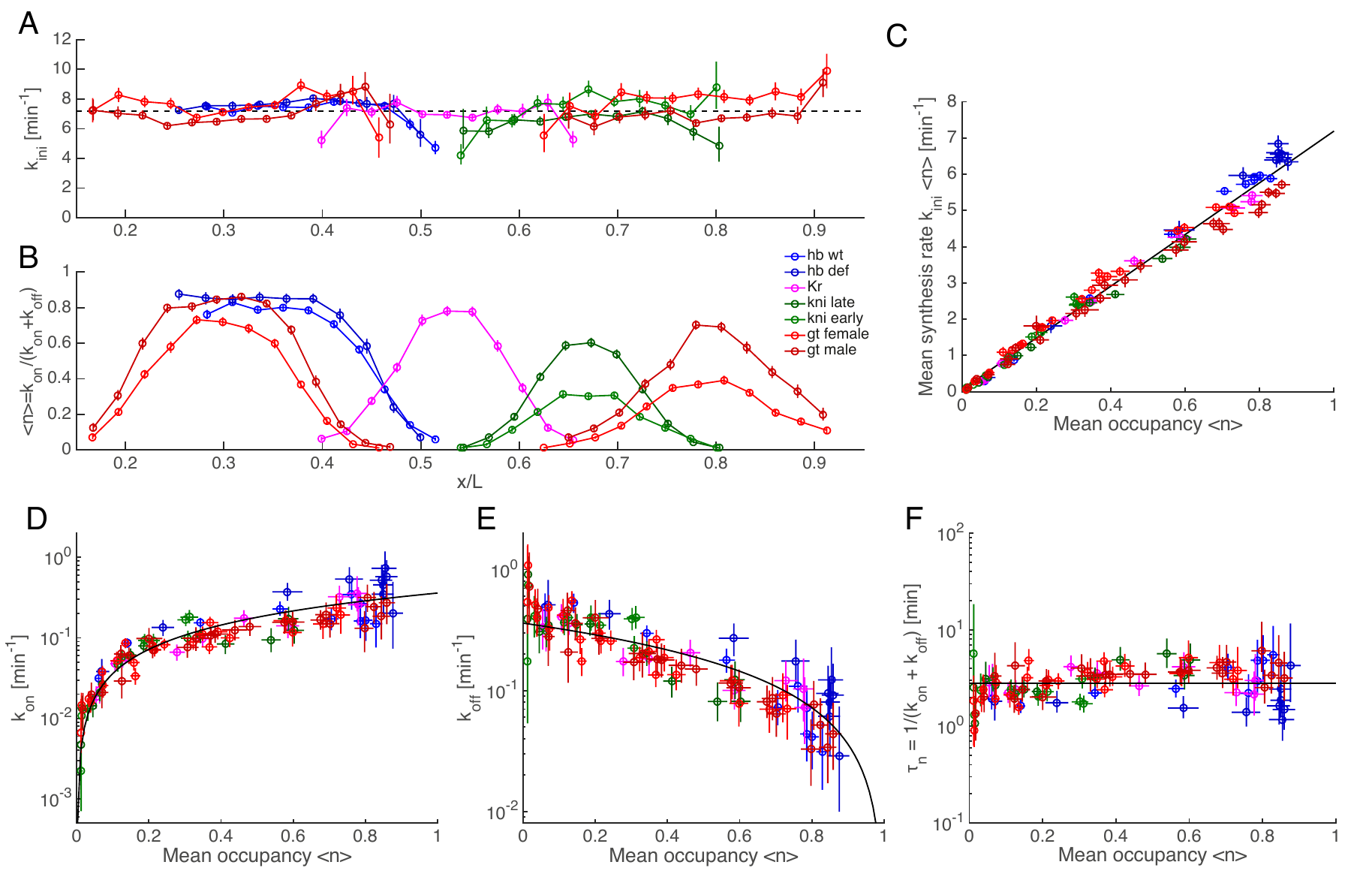}}
\caption{{\bf Transcription rates are tightly constrained across genes.}
A,B) Inferred Pol II initiation rate $k_{\rm ini}$ (A), and promoter mean occupancy $\avg{n}$ (B) based on the two-state model for all trunk gap genes across AP position. 
C) Modulation of transcript mean synthesis rate $k_{\rm ini} \avg{n}$ is fully determined by the mean occupancy $\avg{n}$. 
D,E) Inferred on-rate $k_{\rm on}$ (D) and off-rate $k_{\rm off}$ (E) as a function of the mean occupancy $\avg{n}$ for all gap genes. Solid black line stands for the global trend. 
F) Switching correlation time $\tau_n$ as a function of mean occupancy $\avg{n}$ for all gap genes.
Error bars stand for the 10 to 90$^{\rm th}$ percentiles of the posterior distribution.}
\label{fig:03}
\end{figure*}

Further insight into the molecular mechanisms underlying transcription necessitates knowledge about the absolute scales of the relevant kinetic parameters. Hence, to go beyond the above arguments based on summary statistics, we inferred kinetic rates for each gene and at each position from the full distribution of transcriptional activity while taking into account measurement noise. We thus utilized dual-color smFISH, tagging the 5' and 3' regions of the transcripts with differently colored probe sets that provide two complementary readouts (in C.U.) of the transcription site activity. The measured 5' and 3' activities are correlated via a finite Pol II elongation time (Supplement, Fig. S4). Therefore the possible set of kinetic parameters that could generate the observed activities is constrained (Methods, Figs. S4, S5 and S6). It allows us to deduce the kinetic rate parameters ($k_{\rm ini}$, $k_{\rm on}$, and $k_{\rm off}$) of the two-state model from the joint distribution of 5' and 3' activities for each AP position (Fig. S5). Previous measurements of Pol II elongation rate $k_{\rm elo} = 1.5$ kb/min \citep{Garcia:2013ha} provide an absolute time scale for this system and thus enable inference of endogenous transcription kinetics from chemically crosslinked and otherwise inert embryos.

Strikingly, we uncovered nearly identical modulation behaviors across all expression boundaries, regardless of gene identity or boundary position (Fig. 3). Consistent with our predictions based on summary statistics (Fig. 2), the initiation rate $k_{\rm ini}$ is constant at about 7.2$\pm$1.0 Pol II initiations per minute and does not change across genes or positions (Fig. 3A). Thus during active periods of transcription (i.e. the `ON' state) these genes share the same rate-limiting step(s) in the cascade of molecular interactions leading to productive Pol II elongation. We also observe close agreement between measured and inferred mean activity, as well as good agreement between all other cumulants (Fig. S7). Our inference confirms that all expression boundaries are generated through modulation of the mean promoter occupancy (Fig. 3B). This result also supports the view that the processes that determine $k_{\rm ini}$ are disfavored as mechanisms for controlling overall mRNA synthesis rates. Because these are found to be entirely determined by $\avg{n}$ for all genes, and spanning a similar dynamic range for all boundaries (Fig. 3C), we advocate that promoter occupancy represents the key control parameter describing the formation of expression boundaries.

Surprisingly, both $k_{\rm on}$ and $k_{\rm off}$ change as a function of mean occupancy $\avg{n}$ and are tightly constrained for all genes and across all boundaries (Fig. 3D, E). This suggests that some combination of the switching rates is conserved. Indeed, as predicted by the two-state model above, the correlation time of the switching process is also constant ($\tau_n = 3.0\pm1.2$ min) at all positions over the entire expression range for every gene (Fig. 3F). This constancy arises because $k_{\rm on}$ and $k_{\rm off}$ are modulated simultaneously: an increase in overall expression rate is achieved both by increasing the time spent active (lowering $k_{\rm off}$) and by jointly decreasing the time spent inactive (raising $k_{\rm on}$). Mechanistically this result is surprising because it implies that $k_{\rm on}$ and $k_{\rm off}$ are coordinated such that the promoter switching correlation time is constant and that all boundaries emerge from quantitatively identical modulation of switching rates. These findings are unbiased by our methodology as confirmed using synthetic data (Fig. S6). In addition, even two-fold changes in elongation rate leave our conclusions unaffected, aside from a rescaling of the kinetic parameters (Supplement, Fig. S8).

As noted above, boundaries arise through the combined activities of multiple transcription factor inputs, with each boundary generated by a unique combination of inputs. Current models of boundary formation imply that expression rates of target genes emerge from careful tuning of input factor concentrations and DNA binding affinities \citep{Jaeger:2011gp,Briscoe:2015js,Wu2016}. The complexity and diversity of these inputs therefore leads to an intuitive expectation that kinetic switching rates must also differ between genes. This expectation seems all the more reasonable given the fact that many combinations of $k_{\rm on}$ and $k_{\rm off}$ generate the same $\avg{n}$. A constant correlation time implies within the context of the two-state model that all genes at all positions reach steady-state simultaneously. Such global synchronicity across all loci is maintained for all genes at all expression rates. We propose that such synchronicity is important for ensuring precise and reproducible patterning outcomes, and that this requirement constrains the range of attainable or otherwise desirable values of $k_{\rm on}$ and $k_{\rm off}$. Thus, the apparently complex process of regulating gene expression rates is explained by a conceptually simple strategy of universal modulation.

\section{Discussion}

A multitude of molecular processes influence rates of gene expression. However, it is not clear which, if any, interactions might be more likely than others to determine expression rates, either globally across all genes or for single genes in response to modulatory inputs. Nor have most previous studies documented similarities across endogenous genes. We have developed a single molecule method for measuring modulated kinetic parameters of endogenous genes. Surprisingly, all expression boundaries arise from equivalent switching kinetics, regardless of the differences in upstream regulatory elements. Thus, a simple, common strategy for transcriptional modulation emerges from the apparently complex combination of regulatory interactions specific to each gene. This suggests a shared regulatory basis for transcriptional modulation, the nature of which is currently unknown.

These observations raise the question of whether the common transcriptional kinetics carry a functional advantage \citep{Eldar:2010kka}. The precise positioning of cell fates in early embryos requires minimizing variability between nuclei, which is achieved by a combination of long mRNA and protein lifetimes permitting accumulation, and spatial averaging through the syncytial cytoplasm, which all serve to minimize variability in patterned expression \citep{Little:2013dr}. At the level of transcription, noise minimization would be best achieved via modulation of constitutive activity. Simply changing the Pol II initiation rate $k_{\rm ini}$ at a constitutive promoter (always `ON') would generate the theoretical minimal (Poisson) noise at all levels of activity \citep{Sanchez:2013fp}. The fact that constitutive activity is never observed suggests that some constraint prohibits this system from maintaining these genes in a continuously active state, and/or it is not mechanistically straightforward to alter $k_{\rm ini}$. Instead, our observation of constant switching correlation time at all genes and expression levels suggests that this value plays an important role in facilitating robust patterning. The constant correlation time we measure here implies that each gene at all positions attains steady-state in synchrony, and in a timely manner as its value ($\tau_n = 3.0\pm1.2$ min)  is small relative to the duration of interphase ($\sim$15 min in cycle 13). This suggests that the relative production rates are maintained across a boundary during early development. In addition, the short switching time compared to the combined duration of early nuclear cleavage cycles ensures effective temporal averaging of the switching noise by accumulation of stable transcripts. We therefore propose that both expression timing and noise minimization jointly constrain switching kinetics.

Our findings suggest that all regulatory inputs interface upon a universal set of processes to determine the kinetics of the promoter states. However, these results do not address the mechanistic origins of the common switching rates. Unique combinations of transcription factors determine all the boundaries we have examined. It is possible that protein-DNA affinities have been selected to confer the rates we observe. Alternatively, other events such as promoter-enhancer interactions, chromatin modification dynamics, or pausing/release of Pol II, predominantly determine switching rates. Indeed, it is not clear how transient transcription factor interactions, usually on the order of a few seconds, could generate bursts on the order of minutes \citep{Elf2007,Karpova2008,Gebhardt:2013kl,Morisaki:2014ct,Izeddin:2014dza}. Recent evidence suggests that Mediator and TBP binding play a key role in determining bursting kinetics \citep{Tantale:2016fb}. Thus switching kinetics may not be directly determined by transcription factor binding, but by common transcriptional rate-limiting steps related to recruitment and stability of general factors. Interestingly, more ``mechanistic'' extensions of the two-state model could in principle reproduce the modulation of mean activity $\avg{n}$ at constant correlation time with a single varying kinetic parameter resulting from transcription factor titration across boundaries. A possible extension is a 3-state model, describing a two-step reversible activation, which is consistent with enhancer-promoter interaction and establishment of the transcription machinery \citep{Rieckh:2014ic}. Alternatively a model with an additional noise term such as input noise stemming from diffusion of transcription factors \citep{Tkacik:2008ht,Kaizu:2014eb} could explain the apparent dual modulation of the switching rates observed under the 2-state model. It would then only be when these more detailed models are effectively reduced to the 2-state that the surprising dependence of the rates emerges. Distinguishing these models from the simple 2-state model will require live imaging of the endogenous gap genes.

The universal transcriptional parameters of the gap genes highlight a form of complexity reduction: despite the variety of upstream regulatory elements, all expression boundaries result from similar switching kinetics. As discussed above, whether this simplicity results from an underlying molecular simplicity has yet to be determined. Regardless of the mechanistic means by which these similar rates are achieved, this convergence strongly suggests the presence of global constraints that either limit the range of permitted bursting rates and/or minimize transcription variability in the context of the rapidly developing early embryo. Such convergence might indicate the possibility of a path to general rules in transcription regulation. It is now possible to inquire about the breadth of these generalities and whether they apply to the same gene expressed in different cell types, or to the transcriptome as a whole, or even across organisms. The methods we utilize here are applicable in a variety of systems and permit the discovery of the molecular mechanism(s) conferring universal transcription kinetics. 

\section*{Methods}

\subsection*{Fly strains}

Oregon-R (Ore-R) embryos were used as wild-type. Embryos heterozygous for a deficiency spanning \emph{hb} were collected from crosses of heterozygous adults of the strain w$^{1118}$; Df(3R)BSC477/TM6C. Heterozygotes of the \emph{hb} deficiency, as well as wild-type male and female embryos stained for \emph{gt}, were distinguished from siblings by visual inspection of nascent transcription sites. 
 
\subsection*{FISH and imaging}

We modified our smFISH protocol \citep{Little:2013dr} to minimize background and maximize signal. Embryos were crosslinked in 1xPBS containing 16\% paraformaldehyde for 2 minutes before devitellinization as described in \citep{Lecuyer2008}. Embryos were washed four times in methanol, 5 minutes per wash, with gentle rocking at room temperature, followed by an extended 30-60 minute wash in methanol. Fixed embryos were then used immediately for smFISH without intervening storage. Embryos washed three times in 1X PBS, 5 minutes per wash, at room temperature with rocking. Embryos were then washed 3 times in smFISH wash buffer \citep{Little:2013dr}, 10 minutes per wash, at room temperature. During this time, probes diluted in hybridization buffer \citep{Little:2013dr} were preheated to 37C. Hybridization was performed for 1.5 hr at 37C with vigorous mixing every 15 minutes. During hybridization, smFISH wash buffer was preheated to 37C. Embryos were washed four times with large excess volumes of wash buffer for 3-5 minutes per wash, rinsed twice briefly in PBS, stained with DAPI, and mounted in VECTASHIELD (Vector Laboratories; H-1000). Imaging was performed within 48 hr to ensure high quality signal. DNA oligonucleotides complementary to the open reading frames of genes of interest were conjugated to Atto 565 (Sigma-Aldrich; 72464) and Atto 633 (Sigma-Aldrich; 01464) for all two color measurements.

Imaging was performed by laser-scanning confocal microscopy on a Leica SP5 inverted microscope. We used a 63x HCX PL APO CS 1.4 NA oil immersion objective with pixels of 76$\times$76 nm and z spacing of 340 $\mu$m. We typically obtained stacks representing 8 $\mu$m in total axial thickness starting at the embryo surface. The microscope was equipped ``HyD Hybrid Detector'' avalanche photodiodes (APDs) which we utilized in photon counting mode. This is in contrast to our prior approach \citep{Little:2013dr} in which standard photomultiplier tubes (PMTs) were used to collect two separate smFISH image stacks at two different laser intensities: a low power stack for measuring transcription intensities, and a high power stack to distinguish single mRNAs. The use of low-noise photon-counting APDs in place of standard photomultipliers provided sufficient dynamic range to capture high signal transcription sites and to separate relatively dim cytoplasmic single mRNAs from background fluorescence with a single laser power. This also abrogated the need to calibrate the high- and low-power stacks for comparison. The removal of the calibration step provided an additional reduction in measurement error.

\subsection*{Image analysis and calibration in absolute units}

The raw data are processed according to the image analysis pipeline previously developed and described in details in \citep{Little:2011cg,Little:2013dr}. Briefly, raw images are filtered using a Difference-of-Gaussians (DoG) filter to detect spot objects. A master threshold is applied to separate candidate spots from background. True point-like sources of fluorescence are identified, as they appeared on multiple consecutive z-slices ($>3$) at the same location. All candidate particles are then labeled as transcription sites, cytoplasmic transcripts or noise based on global thresholds. The threshold separating cytoplasmic transcripts from noise is defined as the bottom of the valley between the two peaks on the particle intensity distribution. The threshold for transcription sites depends both on intensity and position, as transcription sites cluster in z and are enclosed in nuclei (segmented from DAPI staining). Intensity of transcription sites is obtained by integrating the signal over a fix cylinder volume (V$V_s=\pi \times 0.76^2 \times 3.06$ ${\rm\mu m^3}$, determined from the PSF). 

We calibrated the integrated intensity of transcription sites $F_s$ by first characterizing the relationship between the fluorescence signal and the density of cytoplasmic transcripts. We defined summation volumes in the embryo ($V\simeq3.8\times3.8\times8.5$ ${\rm\mu m^3}$) avoiding region of high tissue deformation and excluding transcription sites location. For each summation volume we counted the number of detected cytoplasmic transcripts and integrated the fluorescence intensity. At low count density, the fluorescence per summation volume $F$ scales linearly with density $D$ \citep{Little:2013dr}. Fitting a simple linear relationship $F=\alpha D+\beta$, where $\beta$ corresponds to background, enables estimation of a scaling factor $\alpha$ to calibrate transcription sites in ``cytoplasmic units'' (C.U.) for each embryo. Namely, the intensity in C.U. is given by $f=(F_s-bV_s)/\alpha$ where $b$ is the background intensity per pixel in each nucleus. We estimated the measurement error by imaging a control gene (\emph{hb}) in two independent channels using an alternating probe configuration. After normalizing the intensities in C.U., both channels correlate with slope close to one. The measurement error was estimated from the residuals of the orthogonal regression (Fig. S1A, Supplement). In the maximally expressed regions, we measure transcriptional activity with an error of 5\% and relate it to absolute units with an uncertainty under 3.5\% (the largest deviation of the slope 0.968$\pm$0.003 from 1). This represents an error reduction by 3- to 4-fold compared to our previous measurements (assuming multiplicative errors; 6\% vs 20\%). We estimated an upper bound on alignment error between different embryos by assuming that all the embryo-to-embryo variability across boundaries results from misalignment (Supplement).

\subsection*{Two-state model of promoter activity}

The transcriptional activity of a single gene copy was modeled as a telegraph process (ON-OFF promoter switch) with transcript initiation occurring as a Poisson process during the `ON' periods \citep{Peccoud:1995ww}. Within the two-state model, the distribution of nascent transcripts on a gene results from random Pol II initiation in the active state coupled with elongation and termination \citep{Senecal:2014dz,Choubey:2015dl,Xu:2016kd}. For simplicity, we combined elongation and termination as an effective process that was modeled as a deterministic progression. In addition, we assumed that all the kinetic rates of the model are constant in time and identical across embryos. The key parameters of the model are the initiation rate $k_{\rm ini}$, the promoter switching rates $k_{\rm on}$ and $k_{\rm off}$, and the elongation time $\tau_e$. At stead-state, the mean number of transcripts $\avg{g}$ and the variance $\sigma_g^2$ for a single gene copy are given by
\[\avg{g}=g_0\avg{n} \quad \text{and} \quad \sigma_g^2=g_0\avg{n}+g_0^2\avg{n}(1-\avg{n})\Phi(\tau_e/\tau_n)\]
where $g_0=k_{\rm ini} \tau_e$ is the mean number of transcripts in a constitutive regime (always `ON'), $\avg{n}=k_{\rm on}/(k_{\rm on}+k_{\rm off})$ the mean promoter activity, $\tau_n=1/(k_{\rm on}+k_{\rm off})$ the switching correlation time, and $\Phi \in [0,1]$ a noise averaging function (Supplement). Higher cumulants can be readily calculated from the master equations (Supplement). Assuming independent gene copies, the mean and the cumulants simply scales with the number of gene copies (either 2 or 4 copies per nucleus). The mean activity $\mu$ in cytoplasmic units is related to $\avg{g}$ by a conversion factor $C$ that takes into account the exact FISH probes location, i.e. $\mu=C N_g\avg{g}$ with $N_g$ the gene copy number (Supplement). Similarly, cumulants of the data are rescaled by conversion factors that ensure proper normalization of the Poisson background (Supplement).

\subsection*{Inferring transcription kinetics of endogenous genes from two color smFISH}

We performed dual-color smFISH tagging the 5' and 3' region of the transcripts with different probe sets (Fig. S4A). After normalization in cytoplasmic units, both channels offer a consistent readout of the mean and the variability (Fig. S4B-C). Assuming constant elongation rate, the probe locations and the gene length predicts the measured 3' to 5' mean activity ratio (Fig. S4D), albeit with small deviations likely stemming from termination (Fig. S4E-F). The two channels enable estimation of the total nascent transcripts (5' channel) and the fractional occupancy of transcripts along the 5' and 3' portions of the gene at each locus (Fig. S5A-B). The 5' and 3' activities are temporally correlated through the elongation process and thus provide additional information about transcription than single channel smFISH (Fig. S5C). Combining measurements from multiple embryos (Fig. S5B-C), we select nuclei at similar positions (bins of 2.5\% egg length) to generate the joint distribution of 5' and 3' activity across AP position bins (Fig. S3D).  We modeled the joint distribution of 5' and 3' activity based on the two-state model and the exact probe location assuming steady-state (Supplement). The modeled distribution enables calculating the likelihood of normalized fluorescence intensities of transcription sites given a set of kinetic parameters. Using a Bayesian approach, we infer the kinetic rate parameters ($k_{\rm ini}$, $k_{\rm on}$, and $k_{\rm off}$) of the two-state model from the joint distribution at each position (Fig. S5D). The inference framework explicitly takes into account measurement noise and the presence of multiple loci (Supplement). The corresponding best-fitting distributions predicted by the model match the data closely (Fig. S5E), and outliers are mainly explained by measurement and binning noise. To validate this approach, we tested the inference on simulated data using a broad range of parameter values and in presence of measurement noise. This demonstrated that our method is capable of discerning differences as small as 20\% in simulated parameters across an arbitrarily large range of parameter values (Fig. S6). 

\section*{Aknowledgement}

We thank W. Bialek, P. Fran\c cois, J. Kinney, M. Levo, M. Mani, F. Naef, T. Sokolowski, G. Tkacik \& E. Wieschaus for insightful discussions and valuable comments regarding the manuscript. B. Zoller was supported by the Swiss National Science Foundation early Postdoc.Mobility fellowship. This study was funded by grants from the National Institutes of Health (U01 EB021239, R01 GM097275).

\bibliographystyle{unsrt}
\bibliography{bibliography.bib}

\clearpage
\widetext
\begin{center}
\textbf{\large Supplement: Diverse spatial expression patterns emerge from \\ common transcription bursting kinetics}
\end{center}

\setcounter{equation}{0}
\setcounter{figure}{0}
\setcounter{table}{0}
\setcounter{page}{1}
\makeatletter
\renewcommand{\theequation}{S\arabic{equation}}
\renewcommand{\thefigure}{S\arabic{figure}}

\section{Supplementary Figures}

\begin{figure}[h]
\begin{center}
	\includegraphics[scale=0.95]{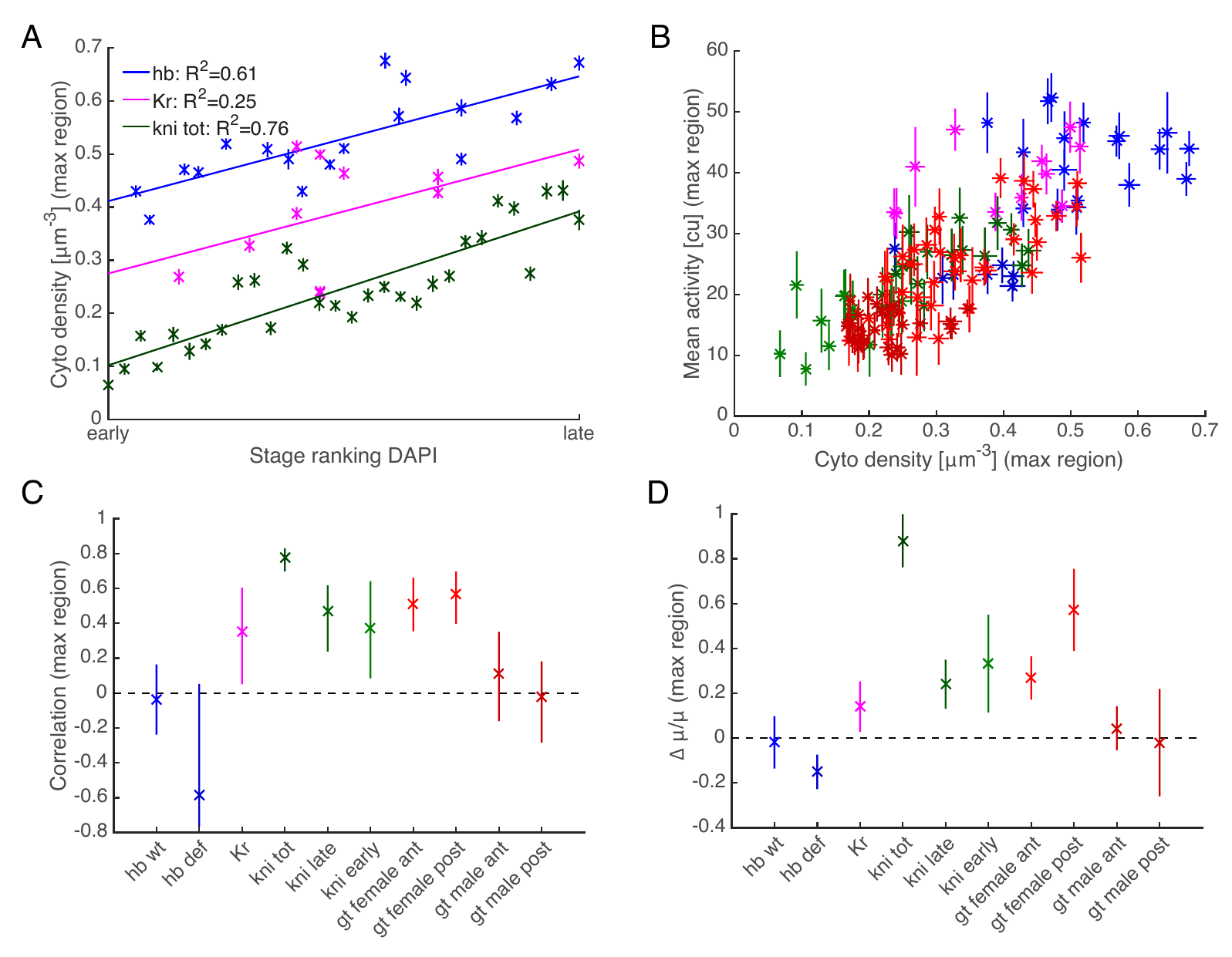}
\end{center}
\caption{\footnotesize Embryos staging. Error bars are given by the 66\% confidence intervals. (A) Cytoplasmic mRNA density as a function of embryo stage (nc13 interphase) estimated from DAPI staining. (B) Mean activity in the maximally expressed regions as a function of the cytoplasmic mRNA density. Each data point corresponds to a single embryo. (C) Pearson correlation coefficient of the mean activity and the cytoplasmic mRNA density calculated over the population of embryos in the maximal expressed regions. (D) Normalized differential activity $\Delta\mu$ explained by timing between early and late embryo as estimated from cytoplasmic mRNA density.}
\label{fig:S0}
\end{figure}

\begin{figure}
\begin{center}
	\includegraphics[scale=0.95]{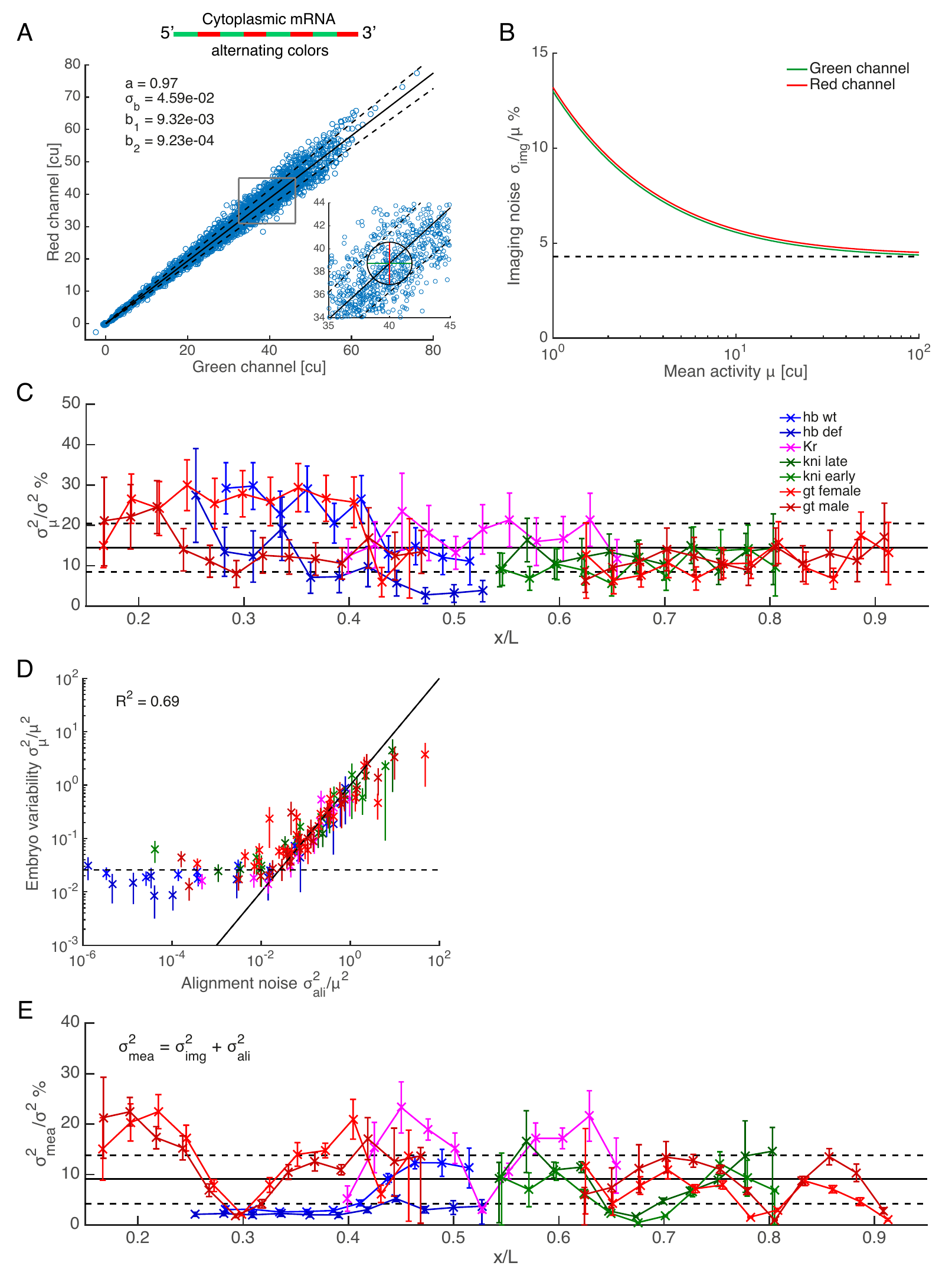}
\end{center}
\caption{\footnotesize Measurement errors and embryos variability. Error bars in C-F are given by the 66\% confidence intervals. (A) Imaging noise model calibrated from dual-color FISH using alternating probe configuration. The data (blue circles) corresponds to the activity of single nuclei measured from 15 \emph{hb} embryos. In absence of measurement noise and without normalization imprecision, both channels should perfectly correlate with slope 1. We characterized the spread along the fitted line (solid line) assuming error on both channel. The dash lines stand for the $1\sigma$ envelope. (B) Imaging noise (CV) as a function of the mean activity for both channels. (C) Fraction of the total variance $\sigma^2$ corresponding to variability of the mean activity across embryos $\sigma_{\mu}^2$ as a function of the AP position. The solid and black dashed lines represent the global mean fraction and the 66\% confidence interval. (D) Embryo variability as a function alignment noise. Each data point corresponds to a single AP bin (2.5\% egg length). The solid line (slope 1) highlights the correlation between the two quantities at the boundaries while the dash line corresponds to the embryo variability in the maximally expressed regions. (E) Fraction of the total variance $\sigma^2$ corresponding to the measurement noise as a function of the AP position. Measurement noise $\sigma_{\rm mea}^2$ is defined as the combination of imaging $\sigma_{\rm img}^2$ and alignement noise $\sigma_{\rm ali}^2$. The solid and black dashed lines represent the global mean fraction and the 66\% confidence interval.}
\label{fig:S1}
\end{figure}

\begin{figure}
\begin{center}
	\includegraphics[scale=0.95]{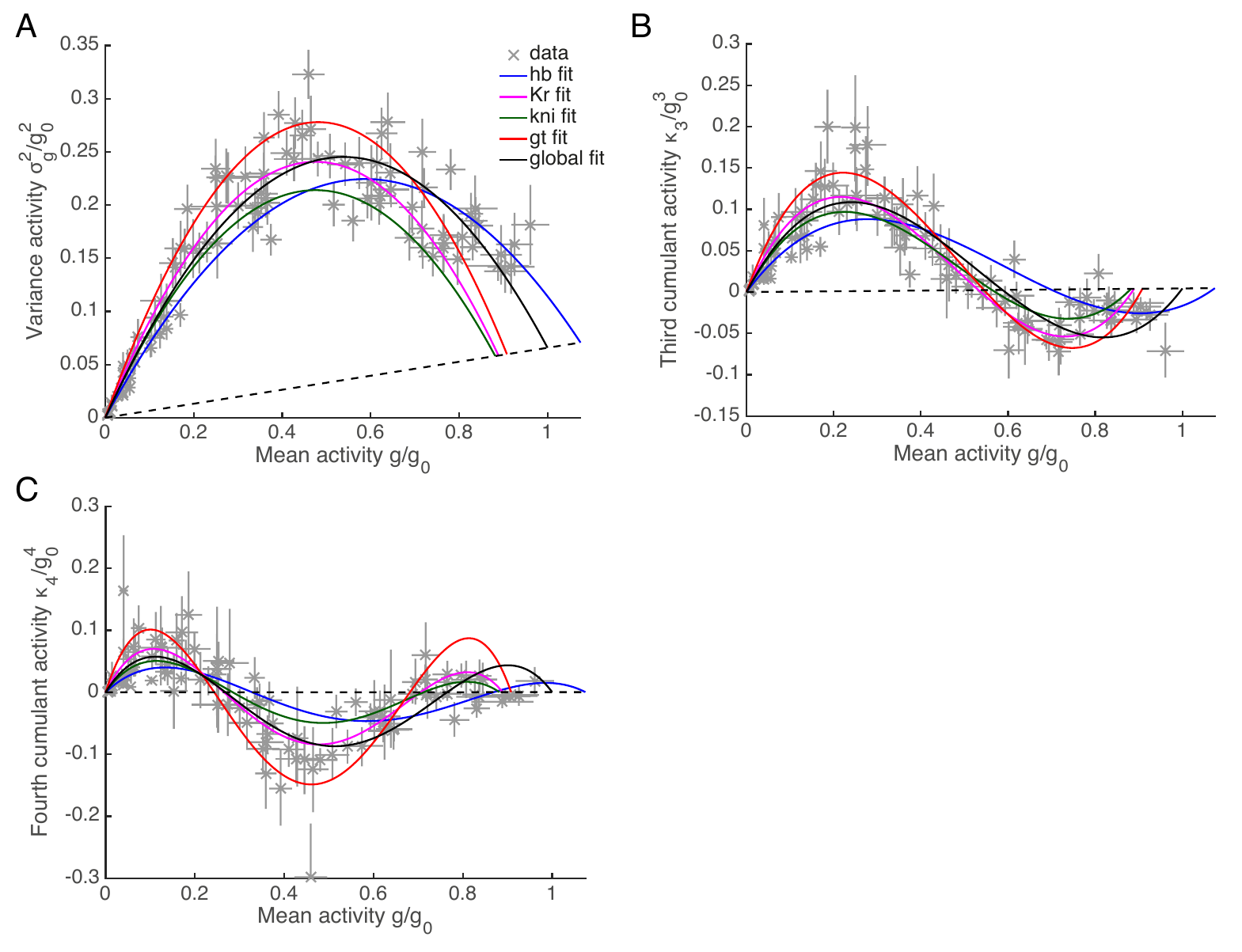}
\end{center}
\caption{\footnotesize Normalized cumulants as a function of normalized mean activity. (A-C) Each data point corresponds to a single AP bin and the error bars are the 66\% confidence intervals. The dash line stands for the Poisson limit. The solid line are $2^{\text{nd}}$ (A), $3^{\text{rd}}$ (B) and $4^{\text{th}}$ (C) order polynomial fits respectively. The fit were performed for each genes independently (color lines) and the black line corresponds to the global fit (Fig 2B).}
\label{fig:S2}
\end{figure}

\begin{figure}
\begin{center}
	\includegraphics[scale=0.95]{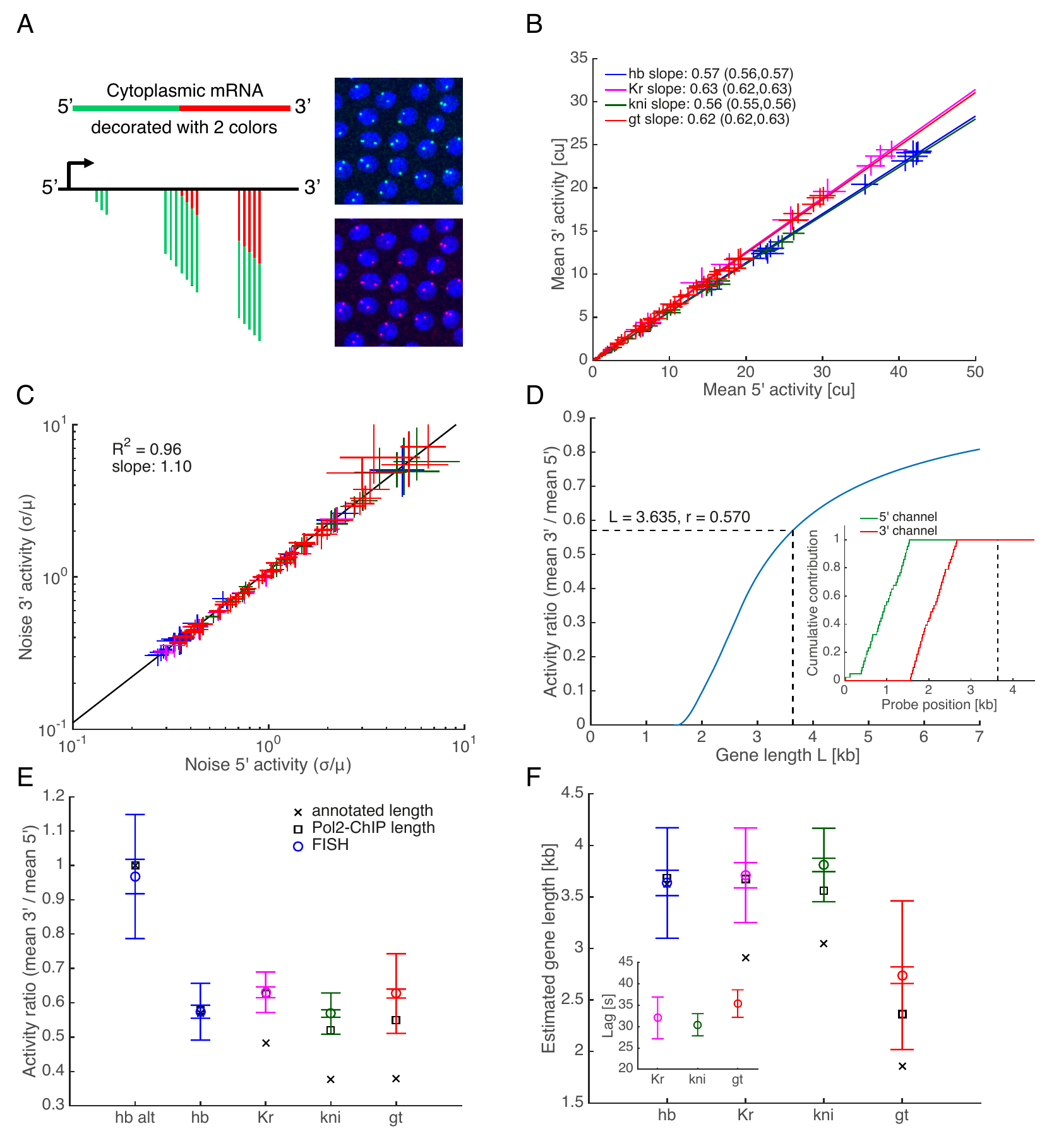}
\end{center}
\caption{\footnotesize Dual-color FISH signal properties link to probe configuration. (A) Schematic of the dual color mRNA-FISH technique. Two independent probe sets hybridized to different fluorophore are designed to target the 5' (green) and 3' region (red) of a transcript of interest. Both channels are physically correlated and provide control over lingering transcripts by quantifying deviation from the expected green-red ratio.(B) Mean 3' versus 5' activity. Each data point correspond to a single AP bin. The slope of the different genes depend on the exact probe configuration. (C) 3' versus 5' noise (CV). (D) Mean activity ratio 3' over 5' as a function of gene length. The probe configuration of \emph{hb} was used as an example. Assuming elongation to occur at constant speed and instantaneous release of transcripts, the ratio is fully determined by the probes' location and the gene length (transcribed region). The activity ratio (blue line) as function of distance results from the ratio of the integrals of the cumulative probe contribution (inset). (E) Activity ratio for each gene. The circles stand for the measured ratio with error bars (2 standard errors and 2 standard deviations respectively) obtained from the propagation of the normalization errors in both channels for all embryos. The crosses correspond to the predicted ratio based on the annotated gene length. The squares are derived from Pol2 occupancy data (Pol2-ChIP). For \emph{Kr}, \emph{kni} and \emph{gt}, Pol2 signal is found a few hundreds bp away from the annotated length suggesting extra processing related to termination. (F) Effective gene length for each gene as determined from the activity ratio. Symbols and error bars similar than in panel E. Assuming an elongation speed of 1.5kb/min, the difference between the effective and annotated gene length can be translated in time (inset). The lag or extra residence time of transcripts at the loci is at most 35 seconds.}
\label{fig:S4}
\end{figure}

\begin{figure}
\begin{center}
	\includegraphics[scale=0.82]{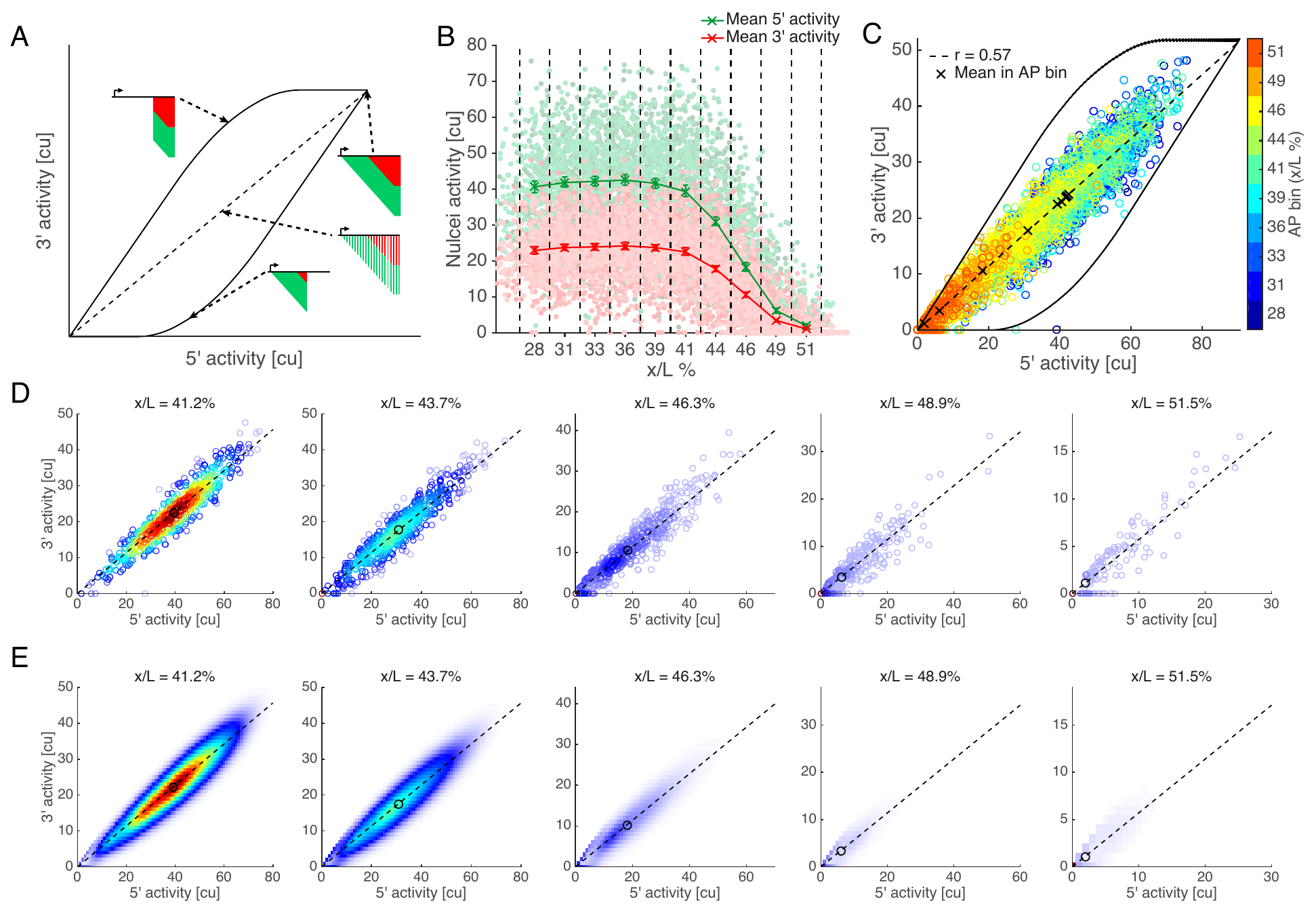}
\end{center}
\caption{\footnotesize Dual color FISH enables inference of transcription dynamics.
(A) The combination of both 5' (green) and 3' (red) readouts constrains the possible configurations of nascent transcript locations and numbers at a given transcription site. This approach offers deeper insight into transcription dynamics compared to conventional single channel mRNA-FISH since both channels are physically correlated.
(B) Transcriptional activity profile for a typical gap gene (\emph{hb}) as a function of AP position for both 5' and 3' channels. A single dot corresponds to the total intensity of nascent transcripts in cytoplasmic unit. As in Figure 1B, 18 embryos have been aligned and overlaid. The vertical dash lines define AP bins covering 2.5\% of egg length; crosses stand for the mean activity across embryo in each bin with error bars corresponding to 66\% confidence intervals.
(C) Dual color measurement space represented as 3' against 5' activity. The solid black line delineates the border of possible measurements given the probe sets configuration, gene length (\emph{hb} 3.6 kb) and the maximal possible Pol2 density (here we assumed a Pol II holoenzyme footprint of 120bp). The dash black line represents the expected ratio of 3' versus 5' activity and defines the subset of equally spaced nascent transcripts configuration. The colored circles stand for the \emph{hb} empirical distributions of 3' versus 5' activity and the color code represents different AP bin; the black crosses correspond to the mean of the distributions for each AP bin and lie on the dash line. For all AP bins, the measurements are enclosed by the envelope of maximal Pol2 density (black line).
(D) Measured distribution of 3' versus 5' activity across AP position for \emph{hb}. The distribution were constructed based on the 2.5\% AP bin defined in Figure 3B. The dash black line represents to the expected ratio of 3' versus 5' activity (r=0.57); the black circle corresponds to the mean of the distribution and lies on the dash line. These distributions are used as input in our inference framework enabling precise inference of the underlying transcriptional kinetics.
(E) Best fitting distribution of 3' versus 5' activity as determined from the empiric distribution in Figure 3D. Of note, the displayed distributions are devoid of measurement noise and represent the theoretical output of the 2-state model given the probe sets configuration and the effective elongation time. Thus, the likelihood of data is essentially the convolution the activity distribution calculated from the 2-state model with the noise measurement distribution that is maximized to determined the kinetic rates $k_{\rm ini}$, $k_{\rm on}$ and $k_{\rm off}$.
}
\label{fig:S3}
\end{figure}

\begin{figure}
\begin{center}
	\includegraphics[scale=0.95]{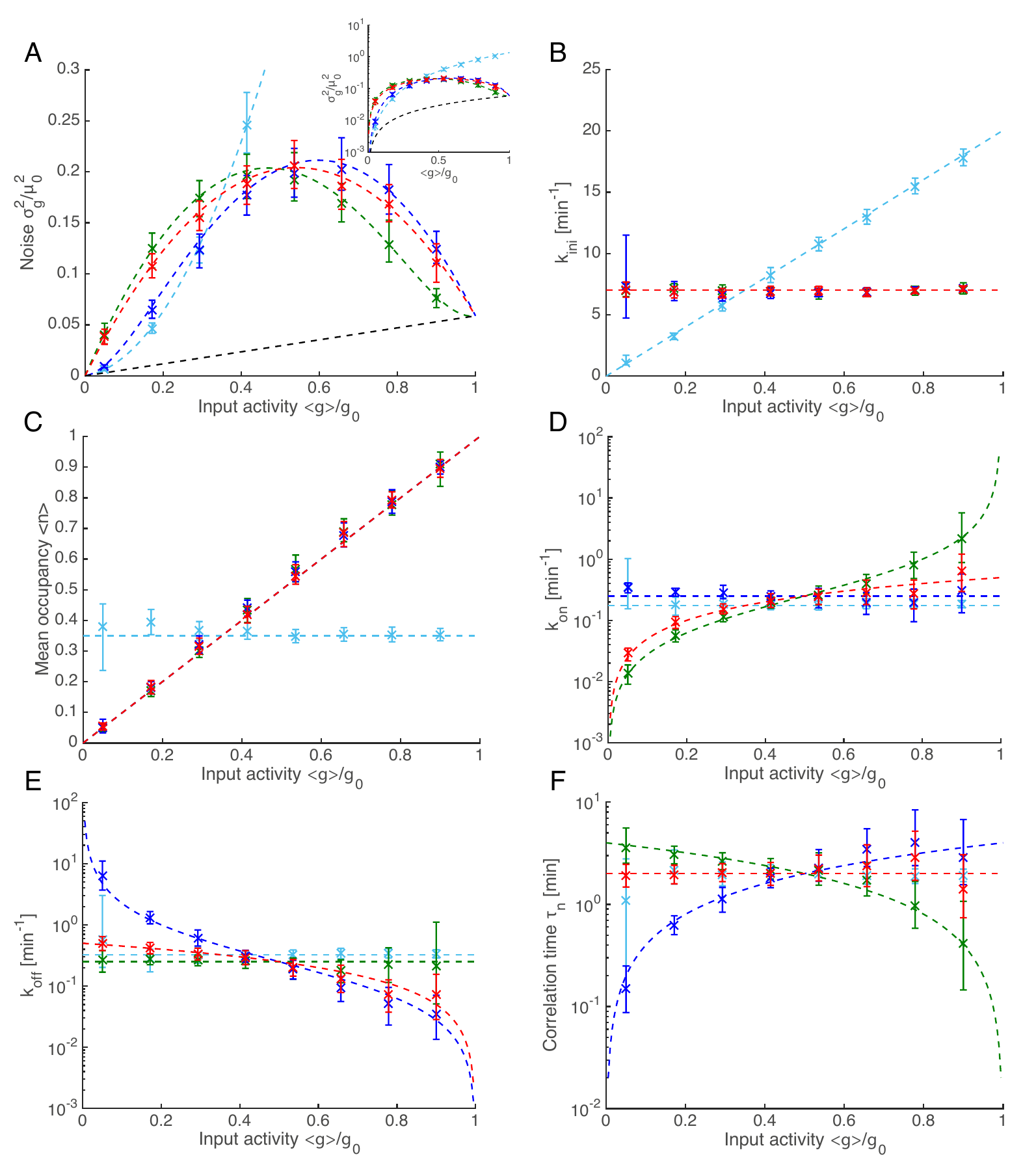}
\end{center}
\caption{\footnotesize Validation of the inference framework for dual-color FISH. We simulated synthetic 3' and 5' nuclei activity data based on four gene copies modeled by a 2-state model, using the probe configuration for \emph{hb} with measurement noise. We tested four different modulation of the mean input activity $\mu$ in the data: 1) initiation rate $k_{\rm ini}$ alone (cyan), 2) on-rate $k_{\rm on}$ alone (green), 3) off-rate $k_{\rm on}$ alone (blue) and modulation of the mean occupancy $\avg{n}$ at constant switching correlation time $\tau_n$ (red). (A-F) The colored crosses stand for the inferred parameters as a function of the input activity. Error bars correspond to the 10 and $90^{\text{th}}$ percentiles of the posterior distribution. The dash lines represents the input (true) parameters used to simulate the data.}
\label{fig:S5}
\end{figure}

\begin{figure}
\begin{center}
	\includegraphics[scale=0.95]{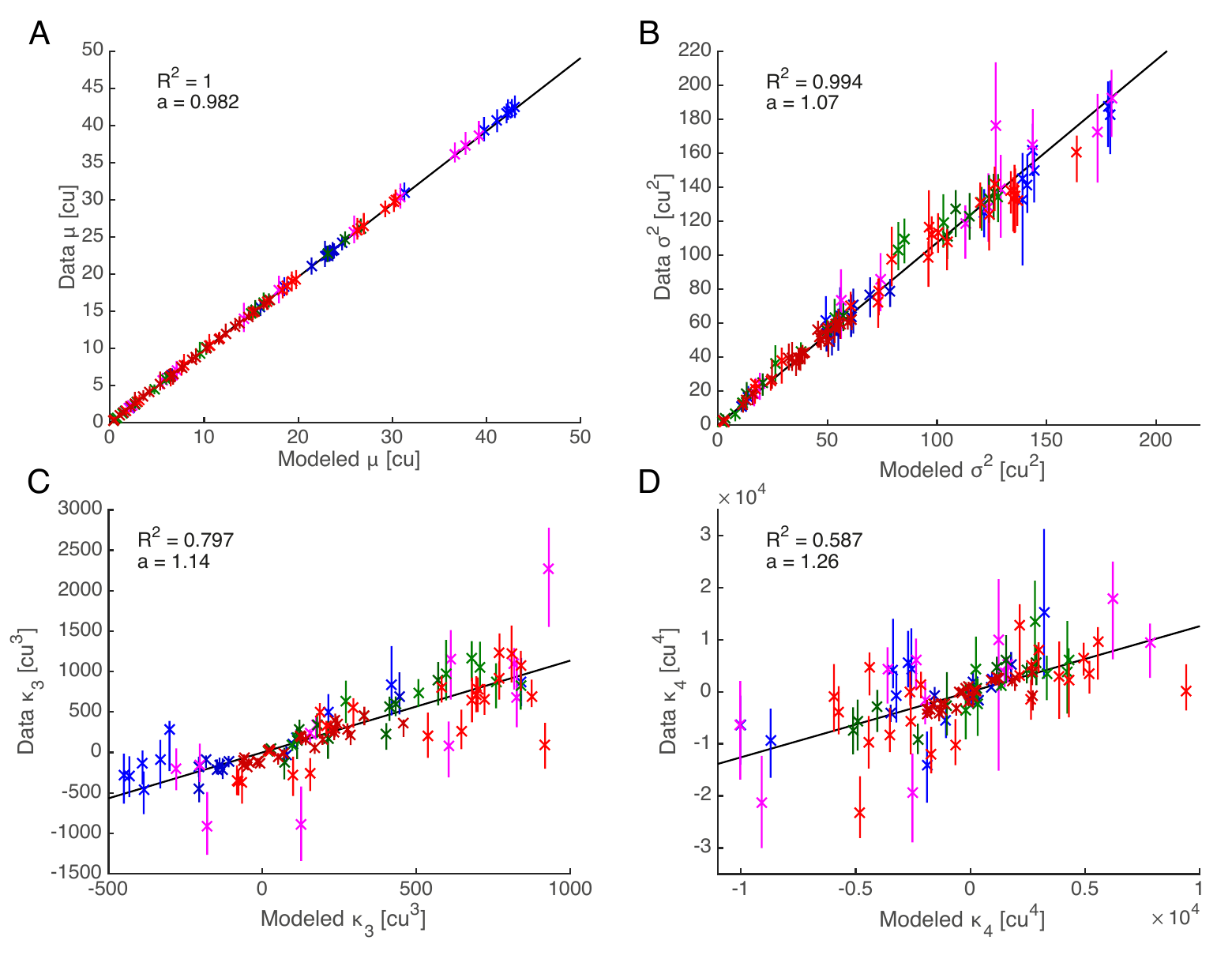}
\end{center}
\caption{\footnotesize Four first cumulants of data (unormalized, in cytoplasmic units) as a function of the ones predicted by the two state-model with best fitting parameters for multiple gene copies ($N_g=2$ or 4). Each data point corresponds to a single AP-bin and the color code stand for the different genes.}
\label{fig:S6}
\end{figure}

\begin{figure}
\begin{center}
	\includegraphics[scale=0.95]{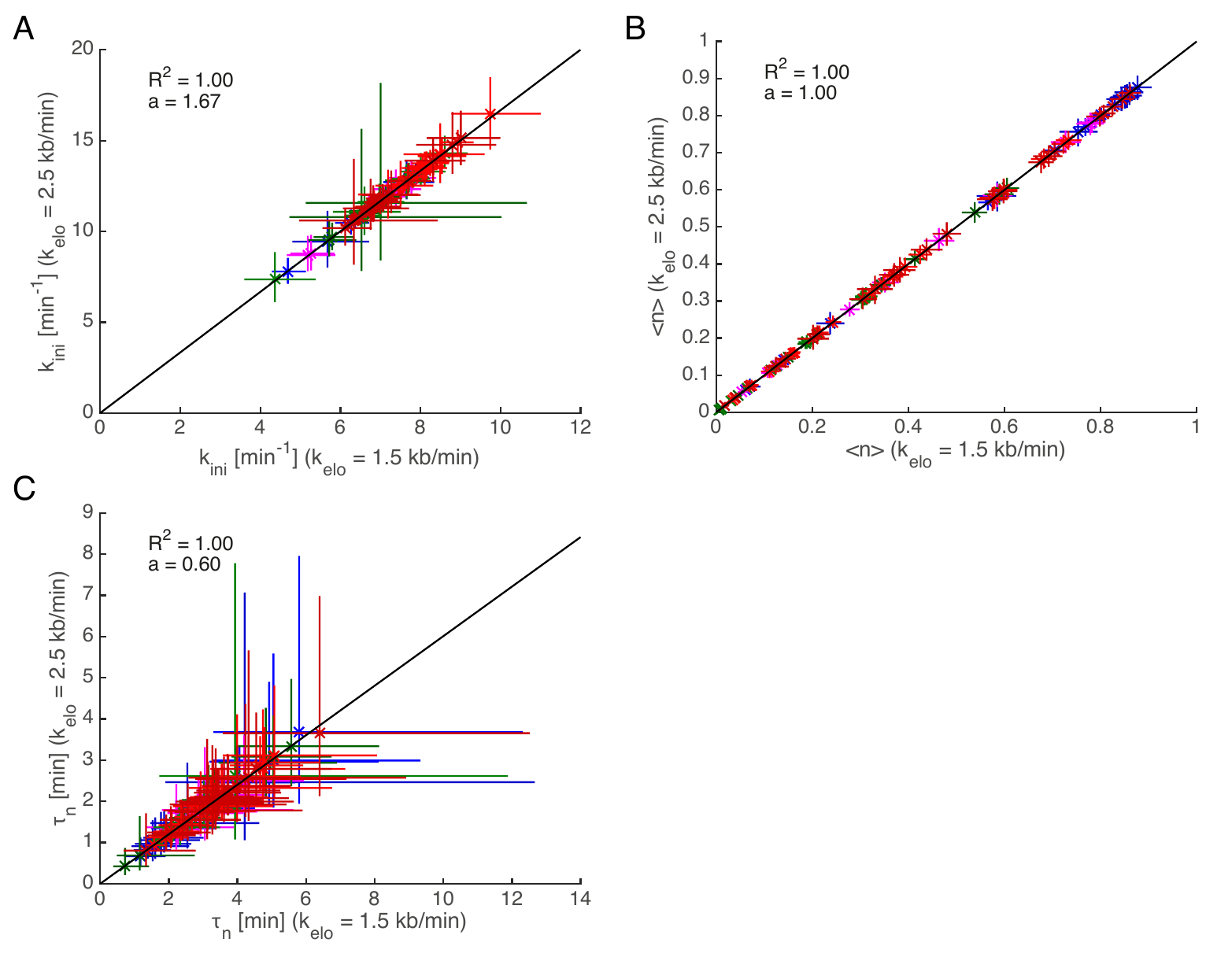}
\end{center}
\caption{\footnotesize Comparison of the inferred transcriptional parameters under different elongation speed $k_{\rm elo}$. Both $k_{\rm ini}$ and $\tau_n$ are rescaled while $\avg{n}$ remains the same.}
\label{fig:S7}
\end{figure}

\clearpage

\renewcommand{\thefigure}{SN\@arabic\c@figure}
\setcounter{figure}{0}

\section{Quantification and measurement error}

\subsection{Embryo staging}

In order to assess the timing of the different embryos, we first manually ranked the different embryos based on timing estimation from DAPI staining. We estimated the interphase stage relying on morphological features of the nuclei (shape and textures) in the DAPI channel. We then verified whether accumulation of cytoplasmic mRNAs correlates with our manual ranking (Fig. \ref{fig:S0}A) . Both approaches lead to similar results and provide a decent proxy for timing. By comparing the average activity of the different embryos in the maximally expressed regions with the cytoplasmic density, we assessed the effect of timing on the mean activity (Fig. \ref{fig:S0}B). We estimated the Pearson correlation coefficient $\rho$ for the different genes and regions (\emph{gt} anterior and posterior regions). Overall, timing explain up to 50\% of the embryo variability in the maximally expressed regions (Fig. \ref{fig:S0}C), with the exception of \emph{kni} that is highly correlated $\rho \sim 0.8$. We thus separated the \emph{kni} embryos in two categories, early and late embryos. We also calculated the difference $\Delta\mu$ in mean activity explained by timing between late and early embryos (Fig. \ref{fig:S0}D), on average the difference is of the order of 5 cu. Although timing affects the activity, the effect is small enough to still warrant the assumption of steady-state.

\subsection{Imaging noise model}

\label{sec:img}

We quantified measurements noise due to imaging and calibration using a two-color smFISH approach, labeling each mRNA in alternating colors along the length of the mRNA. We included 15 $\emph hb$ embryos in the analysis, which corresponds to approximately 4'000 nuclei activity measurements. We then normalized the activity (fluorescence signal) of the nuclei in cytoplasmic units independently in each channel. In absence of noise and provided accurate normalization, both channels would perfectly correlate with slope one. By plotting one channel against the other (Fig. S2A), we assessed the slope and characterized the spread of the data along the expected line.

\label{sec:error}
We build a simple effective model to describe measurement noise:
\begin{equation} 
P(S^{(5)},S^{(3)}|G^{(5)},G^{(3)}) = \mathcal{N}\left(S^{(5)}|\mu=G^{(5)}, \sigma_5^2(G^{(5)})\right) \cdot \mathcal{N}\left(S^{(3)}|\mu=G^{(3)}, \sigma_3^2(G^{(3)})\right)
\label{equ:noise_measure}
\end{equation}
where $S$ stands for the fluorescent signal in cytoplasmic units and $G$ the total nascent transcripts (in C.U.) in absence of noise. We assumed that the measurement errors were normally distributed and independent in both channels, which was motivated by the absence of correlation in the background. We further assumed that the variance would depend on activities, consistent with the increasing spread observed in the data. In order to estimate the variance specific to each channel, we fitted a straight line $y=ax+h$ assuming error on both $x\equiv S^{(5)}$ and $y\equiv S^{(3)}$. We expanded the variance as a function of the scalar projection along the line $v$:
\begin{align}
\sigma^2(v) &= \sigma_b^2 + b_1 v + b_2 v^2 + ... \\
v &= \frac{x+ay}{\sqrt{1+a^2}}
\end{align}
Assuming the same error along $x$ and $y$, we then maximized the following likelihood to estimate the parameters $\theta=\{a,h,\sigma^2_b,b_1,b_2,...\}$:
\begin{equation}
P(\{x_i,y_i\}|\theta) = \prod_{i=1}^{N_d} \frac{1}{\sqrt{(2\pi \sigma^2(v_i))}}\exp{\left(-\frac{(y_i-ax_i-h)^2}{2(1+a^2)\sigma^2(v_i)} \right)}
\end{equation}
Using the Akaike information criterion, we selected the best model which was parametrized by $(a,\sigma^2_b,b_1,b_2)$ with $h=0$ (Fig. S2A). The variances in the noise measurement model (Eq. \ref{equ:noise_measure}) are then given by:
\begin{align}
\sigma_5^2(G) &= \sigma^2(v=\sqrt{G^2 + (a G)^2}) \\
\sigma_3^2(G) &= \sigma^2(v=\sqrt{G^2 + (G/a)^2})
\end{align}
where $\sigma^2(v)=\sigma_b^2 + b_1 v + b_2 v^2$. The resulting imaging noise is shown in Figure S2B.

\subsection{Alignment and embryo variability}

The Anterior-Posterior axis (AP) was determined based on a midsagittal elliptic mask of the embryo in the DAPI channel \citep{Little:2013dr}. Position is obtained by registration of high- and low-magnification DAPI images of the surface. We then fitted constrained splines to approximate the mean activity as a function of the AP position. We used different features of the mean profiles such as maxima and inflection points to refine the alignment between the different embryos. Overall, this realignment procedure enables us to estimate an alignment error of the order of 2\% egg length. 

After alignment, we defined spatial bins along the AP-axis with a width of 2.5\% of egg length. Such a width was a good compromise to balance the sampling and binning error. We next sought to decompose the measured total activity variance $\sigma^2$ into different components related to imaging, alignement, embryo and nuclei variability (Fig. 2E-F). We first estimated the variability of the mean across embryos $\sigma_{\mu}^2$ in each bin (Fig. S2C); we split the total variance $\sigma^2$ in each bin according to the law of total variance:
\begin{equation}
\sigma^2 = \underbrace{\frac{1}{N_e}\sum_{i=1}^{N_e} \sigma_i^2}_{\avg{\sigma^2_{i}}} + \underbrace{\frac{1}{N_e}\sum_{i=1}^{N_e} (\mu_i-\mu)^2}_{\sigma^2_{\mu}}
\end{equation}
where $N_e$ is the total number of embryos and $\mu$ the global mean.

Next we aimed to determine what fraction of $\sigma_{\mu}^2$ is explained by residual misalignment. Assuming that all the variability in the mean at boundaries results from spatial mis-alignment of the different embryos, one can find an upper bound on the residual alignment error $\sigma_x$:
\begin{equation}
\sigma_{\mu}^2 \geq \sigma_{\rm ali}^2 = \left(\frac{\partial \mu}{\partial x}\right)^2 \sigma_{x}^2
\end{equation}
where $\mu$ is the global mean profile as function of AP position $x$. For each gene, we estimated the residual alignment error $\sigma_{x}$ required to explain as much embryo variability as possible (Fig. S2D solid line). Overall we found $\sigma_{\rm x}$ of the order of 1\% egg length. The total embryo variability in the maximally expressed regions cannot be explained by misalignment as $\frac{\partial \mu}{\partial x}\approx 0$ and leads to a noise floor (Fig. S2D dash line). This noise floor can be partly explained by variability in the stage (early vs late interphase) of the different embryos (Fig. \ref{fig:S0}C). In the following we thus split $\sigma_{\mu}^2=\sigma_{\rm ali}^2 + \sigma_{\rm emb}^2$ where $\sigma_{\rm emb}^2$ is the residual embryo to embryo variability.

Finally, we assessed what fraction of the total variance $\sigma^2$ corresponds to combined measurement noise $\sigma_{\rm mea}^2=\sigma_{\rm img}^2 + \sigma_{\rm ali}^2$ where $\sigma_{\rm img}^2$ was estimated in subsection \ref{sec:img}. Total measurement noise $\sigma_{\rm mea}^2$ remains below 25\% of the total variance for all genes and all position (Fig. S2E), and on average reaches $9.1\pm4.9$\%. The remaining variability corresponds to biological variability $\sigma_{\rm bio}^2=\sigma_{\rm nuc}^2+\sigma_{\rm emb}^2$ where  $\sigma_{\rm nuc}^2$ is the nuclei variability and was defined as:
\begin{equation}
\sigma_{\rm nuc}^2 = \sigma^2 - \sigma_{\rm img}^2 - \sigma_{\rm ali}^2 - \sigma_{\rm emb}^2
\end{equation}
Overall, the nuclei variability $\sigma_{\rm nuc}^2$ largely dominates in our data and represents 84\% of the total variance on average (Fig 1E-F).

\section{Telegraph model of transcriptional activity}

\subsection{Temporal evolution of nascent transcripts}

We modeled the transcriptional activity of a single gene copy as a telegraph process (on-off promoter switch) with transcript initiation occurring as a Poisson process during the `on' periods. Elongation of a single transcript was assumed to be deterministic, with fixed speed $k_{\rm elo}$ kb/min. Thus, the elongation time is $\tau_e=L/k_{\rm elo}$ where $L$ is the length of the gene in kb. The key parameters in the model are: the mRNA initiation rate $k_{\rm ini}$, the mRNA elongation time $\tau_{\rm e}$, the on-rate $k_{\rm on}$ and the off-rate $k_{\rm off}$.

The master equation that governs the temporal evolution of nascent transcripts at individual transcription sites is given by
\begin{align}
\frac{d}{dt} P_t(g,n) &=  k_{\rm ini} \delta_{n,1} P_t(g-1,n) - k_{\rm ini} \delta_{n,1} P_t(g,n)  \nonumber \\
&+ k_n P_t(g,n-1) - k_{n+1} P_t(g,n)
\label{equ:master}
\end{align}
with $g$ the number of nascent transcripts (or alternatively the number of Pol II) on the gene and $n$ the promoter state. Here we used the convention that $n=1$ and $n=0$ correspond to the `on' state and `off' state respectively. In addition, the following periodic conditions $n = - 1 \equiv 1$ and $n = 2 \equiv 0$ was used in the equation above.

Of note, we only considered the promoter switching and the initiation of elongation (\ref{equ:master}); we did not explicitly model release of transcripts after termination. The rationale is the following; only the initiation events occurring during the time interval $[t-\tau_e,t]$ contributes to the signal at time $t$, i.e. the elongation time $\tau_e$ determined the ``memory'' of the system. This is correct as long as the release events are instantaneous and termination is fast compared to elongation. Thus, the dynamics of nascent transcripts accumulation on the gene for $t\leq\tau_e$ is obtained by solving the master equation with zero initial transcript on the gene $P_{t_0}(g)=\delta_{g,0}$ and an arbitrary initial distribution of promoter state.

\subsection{Summary statistics}

We can derive the temporal evolutions of the central moments from the master equation (Eq. \ref{equ:master}) \citep{Sanchez:2008jt,Lestas:2008hy}. The means of nascent transcripts $g$ and promoter states $n$ satisfy the following equation:
\begin{equation}
\left\{
\begin{aligned}
\frac{d}{dt} \avg{g(t)} &= k_{\rm ini}  \avg{n(t)} \\
\frac{d}{dt} \avg{n(t)} &= k_{\rm on}  -(k_{\rm on}+k_{\rm off}) \avg{n(t)} \\
\end{aligned}
\right.
\label{equ:mean_temp_evo}
\end{equation}
At steady-state ($\frac{d}{dt} \avg{n}=0$), the mean occupancy of the promoter is simply given by $\avg{n}=k_{\rm on}/(k_{\rm on}+k_{\rm off})$.
Similarly, the covariances satisfy the following set of equations:
\begin{equation}
\left\{
\begin{aligned}
\frac{d}{dt} \sigma_g^2(t) &= 2 k_{\rm ini} \sigma_{gn}(t) + k_{\rm ini} \avg{n(t)} \\
\frac{d}{dt} \sigma_{gn}(t)  &= k_{\rm ini} \sigma_n^2(t)  -(k_{\rm on}+k_{\rm off}) \sigma_{gn}(t) \\
\frac{d}{dt} \sigma_n^2(t) &= -2(k_{\rm on}+k_{\rm off}) \sigma_n^2(t) + k_{\rm on}(1-\avg{n(t)}) + k_{\rm off} \avg{n(t)}
\end{aligned}
\right.
\label{equ:var_temp_evo}
\end{equation}
Assuming zero initial transcripts and promoter at steady-state, one can solve both the mean and variance for $g$. Thus, the initial conditions are given by $\avg{g(t_0)}=0$, $\avg{n(t_0)} = k_{\rm on}/(k_{\rm on}+k_{\rm off})$, $\sigma_g^2(t_0) = 0$, $\sigma_{gn}(t_0) = 0$ and $\sigma_n^2(t_0) = \avg{n(t_0)}(1-\avg{n(t_0)})$. Solving these equations for the elongation time $t=\tau_e$ leads to:
\begin{align}
\avg{g(\tau_e)} &=  g_0 \avg{n} \label{equ:2s_mean} \\
\sigma_g^2(\tau_e) &= g_0 \avg{n} + g_0^2 \avg{n} (1-\avg{n}) \Phi(\tau_e/\tau_n) \label{equ:2s_var}
\end{align}
where $g_0=k_{\rm ini} \tau_e$ is the maximal mean nascent transcript number and $\Phi\in[0,1]$ a noise filtering function that takes into account the fluctuation correlation times. Here, the relevant time scales are the elongation time $\tau_e$ and the promoter switching correlation time $\tau_n=(k_{\rm on}+k_{\rm off})^{-1}$. The variance $\sigma_g^2$ results from the sum of two contributions; the poisson variance $g_0 \avg{n}$ stemming from the stochastic initiation of transcript and the propagation of switching noise:
\begin{equation}
\left(\frac{d\avg{g}}{d\avg{n}}\right)^2 \sigma_n^2 \Phi(\tau_e/\tau_n) = g_0^2 \underbrace{\avg{n} (1-\avg{n})}_{\text{binomial variance}} \Phi(\tau_e/\tau_n)
\label{equ:propagation}
\end{equation}
For deterministic elongation, we find that the noise filtering function is given by:
\begin{equation}
\Phi(x)=2\frac{\exp{(-x)}+x-1}{x^2}
\label{equ:filtering}
\end{equation}
In the limit of fast and slow promoter switching respectively, the noise filtering function reduces to
\begin{align}
\tau_e &\gg \tau_n \quad \lim_{x \rightarrow \infty} \Phi(x) = 0 \\
\tau_e &\ll \tau_n \quad \lim_{x \rightarrow 0} \Phi(x) = 1
\end{align}
Thus, the noise is minimal in the fast switching regime $\tau_e \gg \tau_n$ and reaches the Poisson limit $\sigma_g^2=g_0 \avg{n}$. While in the slow switching regime $\tau_e \ll \tau_n$, none of the switching noise is filtered and the variance is described by a second order polynomial of the mean occupancy $\avg{n}$, i.e. $\sigma_g^2=g_0 \avg{n} + g_0^2 \avg{n} (1-\avg{n})$. Of note, for exponentially distributed life-time of transcripts, such as cytoplasmic mRNA subject to degradation, the results above remain valid except that the noise averaging function becomes $\Phi(x)=(1+x)^{-1}$ with $\tau_e$ the average life-time of the transcripts.

Following a similar approach as in the previous paragraph, higher order moments and cumulants are analytically calculated from the master equations (\ref{equ:master}). The cumulants up to order 3 are equal to the central moments while higher order cumulants can be expressed as linear combination of central moments. The $4^{\text{th}}$ cumulant is given by $\kappa_4 = \mu_4 - 3\mu_2^2$, where $\mu_4$ is the $4^{\text{th}}$ central moment and $\mu_2$ the variance. Assuming promoter at steady-state, we solved the equations for $3^{\text{rd}}$ and $4^{\text{th}}$ moments of $g$ and derive the following analytical expressions for $3^{\text{rd}}$ and $4^{\text{th}}$ cumulants, $\kappa_3$ and $\kappa_4$:
\begin{align}
\kappa_3 &= g_0 \avg{n} + 3 g_0^2 \avg{n} (1 - \avg{n}) \Phi_1(\tau_e/\tau_n) \nonumber \\
&+ g_0^3 \avg{n}(1-\avg{n})(1-2\avg{n}) \Phi_2(\tau_e/\tau_n) \label{equ:k3}\\
\nonumber \\
\kappa_4 &= g_0  \avg{n} + 7 g_0^2 \avg{n} (1 - \avg{n}) \Phi_1(\tau_e/\tau_n) \nonumber \\
&+ 6 g_0^3 \avg{n}(1-\avg{n})(1-2\avg{n}) \Phi_2(\tau_e/\tau_n) \nonumber \\
&+ g_0^4 \avg{n}(1-\avg{n})(\Phi_3(\tau_e/\tau_n)-6\avg{n}(1-\avg{n})\Phi_4(\tau_e/\tau_n)) \label{equ:k4}
\end{align}
where $\Phi_1$,$\Phi_2$,$\Phi_3$ and $\Phi_4$ are noise filtering functions that vanish in the fast switching regime ($\tau_e\gg\tau_n$) and tend to one in the slow switching regime ($\tau_e\ll\tau_n$):
\begin{align}
\Phi_1(x)&=2\frac{\exp{(-x)}+x-1}{x^2} \nonumber \\
\Phi_2(x)&=6\frac{x\exp{(-x)}+2\exp{(-x)}+x-2}{x^3} \nonumber \\
\Phi_3(x)&=12\frac{x^2\exp{(-x)}+4x\exp{(-x)}+6\exp{(-x)}+2x-6}{x^4} \nonumber \\
\Phi_4(x)&=2\frac{\exp{(-x)}^2+4x^2\exp{(-x)}+20x\exp{(-x)}+28\exp{(-x)}+10x-29}{x^4}
\end{align}
The above expressions for the cumulants were tested numerically and are exact. The cumulants are polynomials of the mean promoter activity $\avg{n}$, which follows from the propagation of the binomial cumulants from the switching process. Since the cumulants are extensive, the cumulants for $N_g$ independent gene copies are obtained by multiplying by $N_g$ the expression for a single gene copy (Eq. \ref{equ:k3} and \ref{equ:k4}).

\section{Transcriptional parameters modulation from FISH}

\subsection{Noise mean-relationship in the data}

In the manuscript, we investigated the noise-mean relationship of the transcriptional activity (Fig. 2A). From equations (\ref{equ:2s_var}), we find that the noise (squared coefficient of variation CV$^2$) is given by:
\begin{equation}
\frac{\sigma_g^2}{\avg{g}^2}=\frac{1}{g_0\avg{n}}+\frac{1-\avg{n}}{\avg{n}}\Phi(\tau_e/\tau_n)
\end{equation}
In practice, we measure transcriptional activity in cytoplasmic units (intensity in equivalent number of fully elongated transcripts) and not in Pol II counts $g$ directly. The measured mean activity $\mu$ in cytoplasmic units is proportional to $\avg{g}$, i.e. $\mu = C_1 N_g \avg{g}$ where $C_1\in[0,1]$ is a correction factor accounting for the FISH probe locations on the gene and $N_g$ the number of gene copies (for most genes $N_g=4$, except for \emph{gt} male and \emph{hb} deficient that only have 2 copies). Assuming independence of transcription sites, the measured variance $\sigma^2$ follows a similar relationship, i.e. $\sigma^2 = C_2 N_g \sigma_g^2$ with $C_2\in[0,1]$. The correction factors $C_1$ and $C_2$ are estimated in the next subsection. The measured noise is then given by:
\begin{equation}
\frac{\sigma^2}{\mu^2} = \frac{C_2}{C_1} \left[\frac{1}{\mu} + \frac{1}{C_1 N_g} \frac{\mu_0-\mu}{\mu}\Phi(\tau_e/\tau_n) \right]
\label{equ:noise_mean}
\end{equation}
where $\mu_0 = C_1 N_g g_0$ is the maximal possible expression level in cytoplasmic units. In practice, $C_2/C_1 \approx1$ and $C_1 \approx 0.7$ (Table \ref{tb:correction_factors}, 5' probe location). Modulation of the initiation rate $k_{\rm ini}$ alone would lead to a noise floor at high expression, since the first term corresponding to Poisson noise is inversely proportional to $k_{\rm ini}$ and thus would tend to zero for large $k_{\rm ini}$, whereas the second term does not depend on $k_{\rm ini}$. This behavior is inconsistent with the data and can be ruled out (Fig. 2A). Despite the fact that the $C$ factors and $\tau_e$ are slightly different for each gene, we found that the above noise-mean relationship (Eq. \ref{equ:noise_mean}) captures the overall trend in the data well (Fig. 2A). Interestingly, the derived noise-mean relationship fits the data well when $\Phi$ is assumed constant (Fig. 2A black line, 2 fitting parameters: $\mu_0$ and $\Phi/N_g$), suggesting that the switching correlation time $\tau_n$ does not change much across the range of expression. Although both \emph{gt} male and \emph{hb} deficient follow a similar trend, they deviate from the black line. This is mainly explained by the reduction in gene copies ($N_g=2$ copies instead of 4) that alters the switching noise term (second term) by a factor two and approximately reduces $\mu_0$ by half.

\subsection{Probe configuration and correction factors}

\label{sec:correc}
As we have seen above, the moments of the transcriptional activity in cytoplasmic units are related to the moments in number of nascent transcripts or Pol II counts on the gene by correction factors $C$. Knowing the exact location of the fluorescent probe binding regions along the gene, one can easily calculate the contribution of a single nascent transcript to the signal in C.U. as a function its length $l$:
\begin{equation}
s(l)=\frac{1}{N}\sum_{i=1}^{N}H(l-l_i) = \frac{1}{N} b(l)
\end{equation}
where $H$ is the unit step function, $l_i$ the end position of the $i^{\rm th}$ probe binding region and $N$ the total number of probes. Here, we made the assumption that each fluorescent probe contributes equally to the signal. The number of probes bound to a transcript of length $l$ is given by $b(l)$ and will be denoted $b_i$ for $l\in(l_i,l_{i+1}]$ with $l_{N+1}=L$ the length of a fully elongated transcript. The total fluorescent signal $s$ in cytoplasmic units for $g$ transcripts is given by
\begin{equation}
s = \frac{1}{N}\sum_{i=1}^{N}b_i g_i
\end{equation}
where $g=\sum_{i=1}^{N} g_i$, with $g_i$ the number of transcripts whose length $l$ belongs to the length interval $(l_{i},l_{i+1}]$. Assuming that $g_i$ follows a Poisson distribution with parameter $\lambda_i=k_{\rm ini}\tau_i$ where $\tau_i = (l_{i+1}-l_{i})/k_{\rm elo}$, the mean fluorescent signal $\avg{s}$ is then given by
\begin{equation}
\avg{s} = \frac{1}{N} \sum_{i=1}^{N}b_i \avg{g_i} = \frac{1}{N}\sum_{i=1}^{N}b_i k_{\rm ini}\tau_i = \underbrace{\left( \frac{1}{N} \sum_{i=1}^{N}b_i \tilde{\tau}_i \right)}_{C_1} k_{\rm ini}\tau_e = C_1 \avg{g}
\label{equ:correc_factor}
\end{equation}
where $\tilde{\tau}_i = \tau_i/\tau_e = (l_{i+1}-l_{i})/L$ and $C_1$ the correction factor that relates the mean number of transcripts $\avg{g}$ to the mean fluorescent signal $\avg{s}$ in cytoplasmic units. This relation remains valid for the 2-state model with $\avg{g}=k_{\rm ini}\tau_e\avg{n}$ (Eq. \ref{equ:2s_mean}).

As for the mean, one can calculate the correction factors for the higher moments and cumulants assuming a Poisson background. The second moment is given by
\begin{align}
\avg{s^2} &=\frac{1}{N^2}\avg{\sum_{ij} b_i b_j g_i g_j} = \frac{1}{N^2}\left( \sum_{i \neq j} b_i b_j \avg{g_i} \avg{g_j} + \sum_i b_i^2 \avg{g_i^2} \right)\\
&= \frac{1}{N^2} \left( \sum_{i \neq j} b_i b_j k_{\rm ini}^2\tau_i \tau_j + \sum_i b_i^2 (k_{\rm ini}^2\tau_i^2 + k_{\rm ini}\tau_i) \right) \\
&= \frac{1}{N^2} \left( \sum_{i j} b_i b_j k_{\rm ini}^2\tau_i \tau_j + \sum_i b_i^2 k_{\rm ini}\tau_i \right)
\end{align}
where $\avg{g_i g_j} = \avg{g_i} \avg{g_j}$ since initiation events are assumed independent. This only holds for the Poisson background and is no longer exact for the 2-state model as the switching process would introduce correlations. Nevertheless, the correction factors for the higher moments and cumulants calculated below remain a good approximation under the 2-state model, provided most probes are located in the 5' region. The variance of the signal is finally given by
\begin{equation}
\avg{(\delta s)^2} = \avg{s^2} - \avg{s}^2 = \frac{1}{N^2}\sum_{i=1}^{N} b_i^2 k_{\rm ini}\tau_i = \underbrace{\left( \frac{1}{N^2}\sum_{i=1}^{N} b_i^2 \tilde{\tau}_i \right)}_{C_2} \avg{g} = C_2 \avg{g}
\end{equation}
The calculation above can be generalized to the $3^{\text{rd}}$ and $4^{\text{th}}$ cumulants. We found the following correction factor for the Poisson background:
\begin{equation}
C_k = \frac{1}{N^k}\sum_{i=1}^{N} b_i^k \tilde{\tau}_i \quad \text{for }k=1,...,4
\label{equ:correc_factor_k}
\end{equation}
Calculated values of $C_k$ for each gene and two different configurations of probe locations (5' or 3' region) are given in Table \ref{tb:correction_factors}.

\begin{table}[h!]
\begin{center}
\begin{tabular}{lccc}
\hline
Gene & $k$ & $C_k$ 5' location & $C_k$ 3' location \\
\hline
\emph{hb} & 1 & 0.7431 & 0.4237 \\
& 2 & 0.6806 & 0.3705 \\
& 3 & 0.6510 & 0.3433 \\
& 4 & 0.6337 & 0.3268 \\
\hline
\emph{Kr} & 1 & 0.7447 & 0.4615 \\
& 2 & 0.6804 & 0.4157 \\
& 3 & 0.6547 & 0.3905 \\
& 4 & 0.6408 & 0.3739 \\
\hline
\emph{kni} & 1 & 0.6368 & 0.3507 \\
& 2 & 0.5547 & 0.3116 \\
& 3 & 0.5264 & 0.2912 \\
& 4 & 0.5127 & 0.2785 \\
\hline
\emph{gt} & 1 & 0.8072 & 0.5087 \\
& 2 & 0.7509 & 0.4616 \\
& 3 & 0.7246 & 0.4383 \\
& 4 & 0.7093 & 0.4240 \\
\hline
\end{tabular}
\end{center}
\caption{Conversion factors used to convert the $k^{\rm th}$ cumulants from cytoplasmic units to Pol II counts or number of transcripts for each gene and two different configurations of probe locations.}
\label{tb:correction_factors}
\end{table}

\subsection{Cumulants for a single gene copy}

In the manuscript, we calculated the $2^{\text{nd}}$, $3^{\text{rd}}$ and $4^{\text{th}}$ cumulants from the data (Fig. 2B-D) and compared them with the ones predicted by the 2-state model under different kinetic modulation schemes (Fig. 2G-I). For independent random variables, the cumulants have the property to be extensive, which is convenient as the measured transcriptional activities result from the sum of 2 to 4 gene copies. 

We first converted the $k^{\text{th}}$ cumulants $\tilde{\kappa}_k$ computed from the data in cytoplasmic units to Pol II counts (or number of nascent transcripts) for a single gene copy with a normalized gene length:
\begin{equation}
\kappa_k = \frac{1}{C_k N_g} \left(\frac{\avg{L}}{L}\right)^k \tilde{\kappa}_k
\end{equation}
where $\kappa_k$ is the $k^{\text{th}}$ cumulant in Pol II counts for a single gene copy, $L$ the gene length, $N_g$ the number of gene copies (4 for most genes, except \emph{gt} male and \emph{hb} deficient that only have 2 copies) and $C_k$ a correction factor for the $k^{\text{th}}$ cumulant to ensure proper normalization of the Poisson background (Eq. \ref{equ:correc_factor_k} and Table \ref{tb:correction_factors}). The annotated gene length $L$ varies between 1.8 to 3.6 kb for the ``gap'' genes. In the following we used an effective gene length that is slightly larger and takes into account the possible lingering of fully elongated transcripts at the loci (Table \ref{tb:gene_length}). This effective gene length can be estimated from the dual color FISH data (cf. Section \ref{sec:effective_gl}). 

We then fitted a second order polynomial of the mean activity $\avg{g}$ to the variance $\sigma_g^2$ (Fig. 2B) in order to estimate the maximal activity $g_0$, which was defined as the crossing point between the Poisson background (Fig. 2B dash line) and the fitted variance (solid line). Similarly, we fitted $3^{\text{rd}}$ and $4^{\text{th}}$ order polynomial of the mean activity to the cumulants $\kappa_3$ and $\kappa_4$ (Fig. 2C-D), constrained to reach the Poisson limit at $g_0$. Of note, the cumulants of the Poisson distribution are all equal to the mean. As can be seen in Figure 2B-D, the polynomial fits (solid lines) capture the main trend observed in the data, suggesting a simple relationship between the cumulants and the mean. It follows that the underlying activity distribution is essentially a universal single parameter distribution whose parameter is the mean activity. To test the extent of the universality, we repeated the analysis above of each gap gene individually (Fig. \ref{fig:S2}A-C). The individual fits (colored solid lines) remain relatively close to each other. Although the fits for \emph{hb} slightly deviate from the other genes, the global shape of the cumulants is conserved.

We then tested whether the 2-state model could qualitatively explain the universal trend observed in the cumulants. We investigated various single transcriptional parameter modulations, assuming a fixed elongation time $\tau_e = \avg{L}/k_{\rm elo}$ (Fig. 2F-I). Consistent with the initial observation based on the noise-mean relationship (Fig. 2A), modulation of the mean occupancy $\avg{n}$ with constant switching correlation time $\tau_n$ leads to good qualitative agreement with the data (Fig. 2F-I), particularly when $\tau_n \geq \tau_e$ (red line). Notably, modulation of $k_{\rm on}$ (cyan line) or $k_{\rm off}$ (green line) alone do not capture the shape of the cumulants very well.

\section{Transcriptional parameters from dual-color FISH}

\subsection{Dual-color FISH and effective gene length}

\label{sec:effective_gl}

In the manuscript, we aimed to learn the kinetic parameters of the 2-state from the FISH data for each gene along the AP-axis. In order to constrain the possible set of kinetic parameters that could generate the observed activity levels, we performed dual-color smFISH tagging the 5' and 3' region of the transcripts with different probe sets (Fig. \ref{fig:S4}A). The mean (Fig. \ref{fig:S4}B) and the noise (CV, Fig \ref{fig:S4}C) of the 5' and 3' channel strongly correlates as expected from the experimental design. Thus, the two channels offer a consistent readout of the variability.

For each gene, given the 5' and 3' FISH probe configurations, we calculated the expected ratio of 3' over 5' signal $r=C_1^{(3)}/C_1^{(5)}$ according to Eq. \ref{equ:correc_factor} (Fig. \ref{fig:S4}D). The predicted ratio is overall smaller than the measured one (Fig. \ref{fig:S4}E). It suggests that nascent transcripts might be retained at transcription sites for a short duration. We then calculated for each gene, the effective length that would be consistent with the measured ratio (Fig. \ref{fig:S4}F, Table \ref{tb:gene_length}). Assuming an elongation rate $k_{\rm elo}=1.5$kb/min \citep{Garcia:2013ha}, we estimated the lag consistent with the length difference between the effective and annotated length (Fig. \ref{fig:S4}F inset). Nascent transcripts remain at the loci for at most 35s.

\begin{table}[h!]
\begin{center}
\begin{tabular}{lccc}
\hline
Gene & Annotated length [bp] & Pol II ChIP [bp] & Effective length [bp] \\
\hline
\emph{hb} & 3621 & 3685 & 3635 \\
\emph{Kr} & 2909 & 3671 & 3710 \\
\emph{kni} & 3049 & 3561 & 3810 \\
\emph{gt} & 1855 & 2360 & 2740 \\
\hline
\end{tabular}
\end{center}
\caption{Gene length of the `gap' genes. The annotated gene length was obtained from UCSC Genome Browser (BDGP Release 6 + ISO1 MT/dm6) Assembly. The length estimated from Pol II ChIP density profiles is longer overall, suggesting that Pol II might be transcribing a few hundreds base pairs downstream during termination. We estimated an effective gene length from the ratio of 5' and 3' activity measurements assuming a constant elongation rate. Overall, the effective length is larger than the annotated one and consistent with the Pol II profiles, suggesting that transcripts might be retained at the loci for a short amount of time.}
\label{tb:gene_length}
\end{table}

\subsection{Distribution of nascent transcripts}

To learn transcriptional parameters, we designed an inference framework based on the 2-state model. Two key distributions in this framework are the steady-state distribution of nascent transcripts (or Pol II number) on the gene and the propagator that describes the temporal evolution of an arbitrary distribution of nascent transcripts. Both distributions can be derived from the master equation (\ref{equ:master}). Although the master equation can be solved using generating functions \citep{Walczak:2012id,Xu:2016kd}, we followed another route that can be easily extended to multi-state system and remains computationally tractable. The master equation can be written in terms of a Lagrangian operator $A$ containing the propensity functions of the different reactions:
\begin{equation}
\frac{d}{dt} P_t(g,n) = A P_t(g,n)
\label{equ:master2}
\end{equation}
\paragraph{Single gene copy} After appropriate truncation on the transcript number (setting an upper bound for the maximum number of nascent transcripts) \citep{Munsky:2006es}, the $A$ operator can be written in terms of a sum of tensor products of different matrices:
\begin{equation}
A = I_G \otimes N_2 + K_G \otimes R_2
\label{equ:lagrangian_operator}
\end{equation}
with $I_G$ standing for the identity matrix of size $G+1$ where $G$ is the maximum number of transcripts after truncation. The matrix $N_2$ encodes the rates of the possible transitions for the 2-state promoter and is given as follows:
\begin{equation}
N_2= \begin{pmatrix}
  -k_{\rm on} & k_{\rm off}\\
  k_{\rm on} & -k_{\rm off}\\
 \end{pmatrix}
 \label{equ:switch_operator}
\end{equation}
while $K_G$ describes the initiation of transcripts
\begin{equation}
K_G = \begin{pmatrix}
  -k_{\rm ini} & 0 & \hdots & \hdots & 0 \\
  k_{\rm ini} & -k_{\rm ini}  & \ddots & \ddots  & \vdots \\
  0 & k_{\rm ini} & \ddots & \ddots & \vdots \\
  \vdots & \ddots & \ddots & -k_{\rm ini} & 0 \\
 0 & \hdots & 0  & k_{\rm ini} & -k_{\rm ini}
 \end{pmatrix}
\end{equation}
and $R_G$ indicates in which promoter state initiation occurs, $R_G(n,n') = 1$ if $n=n'=1$ and zero otherwise. The propagator of the resulting finite system (Eq. \ref{equ:master2}) can be expressed as a matrix exponential of the $A$ operator:
\begin{equation}
P_t(g,n | g',n') = \exp{\left( A t \right)}
\label{equ:full_prop}
\end{equation}
The distribution of nascent transcripts $P(g)$ for a gene of length $L$ is calculated using the propagator above with $t=\tau_e \equiv L/k_{\rm elo}$ and the initial conditions. With initially zero nascent transcript, $P(g)$ is then given by:
\begin{equation}
P(g) = \sum_{n,g',n'} P_{\tau_e}(g,n | g',n') \delta_{g',0} P(n')
\label{equ:nascent_distrib}
\end{equation}
where $P(n)$ specifies the initial distribution of promoter state. The distribution $P(g)$ can be computed efficiently by directly estimating the action of the initial vector on the matrix exponential \citep{Sidje:1998uq}. Assuming the promoter at steady-state, $P(n)$ is then given by:
\begin{equation}
P(n) = \left\{\begin{aligned}
\avg{n} \quad\quad &\text{for} \quad n=1\\
1-\avg{n} \quad &\text{for} \quad n=0
\end{aligned}
\right.
\end{equation}
with $\avg{n}=k_{\rm on}/(k_{\rm on}+k_{\rm off})$.

\paragraph{Multiple independent gene copies}

\begin{figure}[h]
\begin{center}
	\includegraphics[scale=0.8]{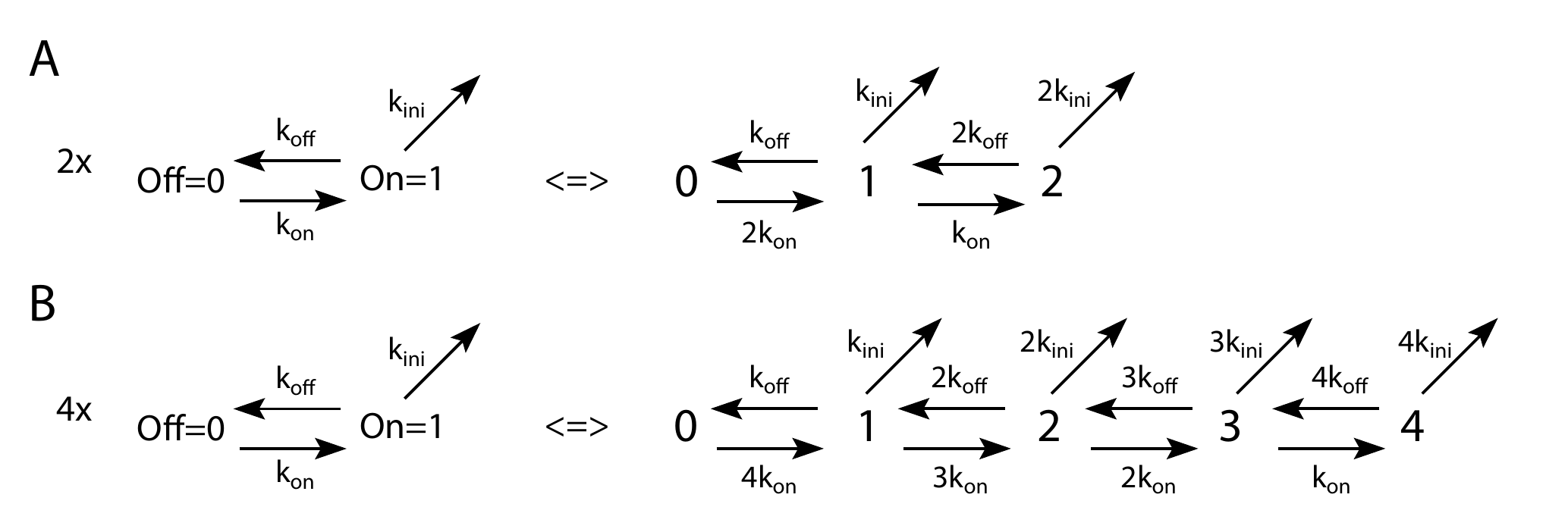}
\end{center}
\caption{Equivalent representations of the superposition of two (A) and four (B) independent and identical telegraph processes.}
\label{fig:model}
\end{figure}

Provided each gene copy is independent and undistinguishable, the combination of two and four gene copies can be represented by a three- and five-state promoter model (Fig. \ref{fig:model}). The corresponding $N$ and $R$ matrices are given by:

\begin{equation}
N_3= \begin{pmatrix}
  -2 k_{\rm on} & k_{\rm off} & 0\\
  2 k_{\rm on} & -(k_{\rm off}+k_{\rm on}) & 2 k_{\rm off} \\
  0 & k_{\rm on} & -2 k_{\rm off} \\
 \end{pmatrix}
 \label{equ:switch_operator_2Al}
\end{equation}

\begin{equation}
R_3= \begin{pmatrix}
  0 & 0 & 0\\
  0 & 1 & 0 \\
  0 & 0 & 2\\
 \end{pmatrix}
 \label{equ:r_operator_2Al}
\end{equation}

\begin{equation}
N_5= \begin{pmatrix}
  -4 k_{\rm on} & k_{\rm off} & 0 & 0 & 0\\
  4 k_{\rm on} & -(k_{\rm off}+3 k_{\rm on}) & 2 k_{\rm off} & 0 & 0 \\
  0 & 3 k_{\rm on} & -2 (k_{\rm off} + k_{\rm on}) & 3 k_{\rm off} & 0 \\
  0 & 0 & 2 k_{\rm on} & -(3 k_{\rm off}+k_{\rm on}) & 4 k_{\rm off} \\
  0 & 0 & 0 & k_{\rm on} & -4 k_{\rm off} \\
 \end{pmatrix}
 \label{equ:switch_operator_4Al}
\end{equation}

\begin{equation}
R_5= \begin{pmatrix}
  0 & 0 & 0 & 0 & 0\\
  0 & 1 & 0 & 0 & 0\\
  0 & 0 & 2 & 0 & 0\\
  0 & 0 & 0 & 3 & 0\\
  0 & 0 & 0 & 0 & 4\\
 \end{pmatrix}
 \label{equ:r_operator_4Al}
\end{equation}

The distribution of nascent transcripts is calculated according to Eq. (\ref{equ:nascent_distrib}), with the propagator $P_t(g,n | g'n')$ computed from the updated $A$ operator (Eq. \ref{equ:lagrangian_operator} \& \ref{equ:full_prop}). The steady-state distribution of the $N$-gene copy system is given by:
\begin{equation}
P(n) = \begin{pmatrix} N \\ n  \end{pmatrix} \avg{n}^n (1-\avg{n})^{N-n} \quad \text{with $n\in\{0,1,...,N\}$}
\label{equ:steady_state_multi}
\end{equation}
where $\avg{n}=k_{\rm on}/(k_{\rm on}+k_{\rm off})$ is the steady-state occupancy of a single promoter.

\subsection{Joint distribution of 5' and 3' activity}

In the following subsection, we lay out the approach used to calculate the joint distribution of 5' and 3' activity for an arbitrary configuration of 5' and 3' FISH probes. Analytic solutions for steady-state distributions with idealistic single color probe configuration exist \citep{Xu:2016kd}, but solutions for arbitrary probe configurations and multi-color FISH are cumbersome. Here, the computational approach is general enough and can be applied to a large class of transcription model, at or out of steady-state (transient relaxation), provided the elongation process is assumed deterministic.

\begin{figure}[h]
\begin{center}
	\includegraphics[scale=0.8]{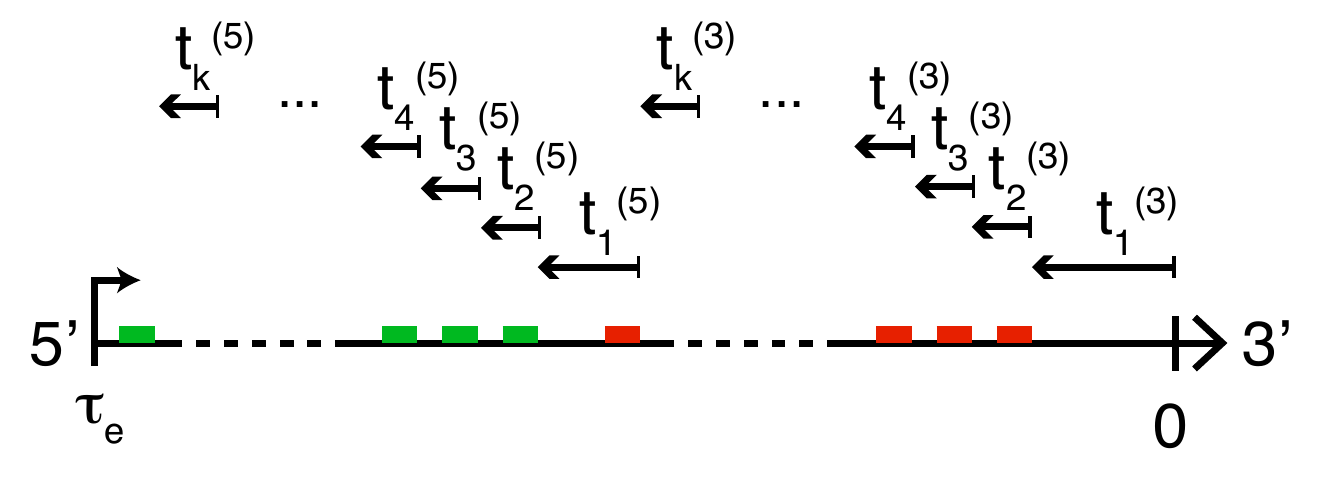}
\end{center}
\caption{Temporal hierarchy of dual color probe configuration.}
\label{fig:hierarchy}
\end{figure}

The measured 5' and 3' transcriptional activities result from partially elongated nascent transcripts. Each fluorescent probe is assumed to be instantaneously bound and to contribute equally to the total fluorescence. Thus, the fluorescent signal of each nascent transcript is proportional to the number of probe binding regions that have been transcribed. In order to calculate the joint distribution, one needs to proceed backward in time. Starting from the 3' end up to the 5' end of the gene, we accumulate the contribution of nascent transcripts that could have been initiated in the interval separating two successive probe regions. Since we assumed elongation to occur at constant speed, the distance interval between two successive probe regions can be converted into a time interval. Doing so for each distance interval leads to the following temporal hierarchy (Fig. \ref{fig:hierarchy}). For instance, only transcripts initiated during the time interval $[-\tau_e,-\tau_e+t_1^{(3)}]$ fully contribute to the 3' (red) signal measured at time $t=0$. The probability to initiate $g$ nascent transcripts during this time interval is given by the propagator $P_{t_1^{(3)}}(g, n|g', n')$ (Eq. \ref{equ:full_prop}), where $n$ is the promoter state.

For any model of promoter activity that only consider the stochastic initiation of transcripts (as a Poisson process) and deterministic elongation with instantaneous release, the propagator will satisfy the following equality:
\begin{equation} 
P_{t}(g,n|g',n') = P_{t}(g-g',n|0,n')
\end{equation}
Thus, one only needs to calculate $P_{t}(g,n|0,n')\equiv P_{t}(g,n|n')$, which can be computed much faster than the matrix exponential (Eq. \ref{equ:full_prop}) \citep{Sidje:1998uq}. It then follows that the Chapman-Kolmogorov equation for the time propagation reduces to a discrete convolution:
\begin{equation}
P_{t_2+t_1}(g_2,n_2|n_0)=\sum_{n_1}\sum_{g_1=0}^{g_2} P_{t_2}(g_2-g_1,n_2|n_1)P_{t_1}(g_1,n_1|n_0)
\end{equation}
This property is used extensively in the following calculation of the joint distribution.

The computation of the joint distribution is performed according to a dynamic programming approach that can in principle be applied to an arbitrary number of color probes. We first calculate recursively the 3' contribution (red probes) to the signal $P^{(3)}(\tilde{G}_k,G_k,n_k)$, where $\tilde{G}_k$ stands for the total signal in probe space, $G_k$ the total number of nascent transcripts, $n_k$ the promoter state and $k$ the total number of probes covering the 3' region. We then calculate the 5' contribution in a similar fashion, $P^{(5)}(\tilde{G}_k|n_0)$. Lastly, we combine both components to generate the final joint distribution $P(\tilde{G}^{(5)},\tilde{G}^{(3)})$ in probe space.

\paragraph{Step 1: calculate the 3' contribution}

The initial distribution is given by:
\begin{align}
P^{(3)}(\tilde{G}_2,G_2,n_2) &\equiv P^{(3)}(\underbrace{(k-1)g_2+kg_1}_{\tilde{G}_2},\underbrace{g_2+g_1}_{G_2},n_2) \nonumber \\
&=\sum_{n_0,n_1} P_{t_2^{(3)}}(g_2,n_2|n_1) P_{t_1^{(3)}}(g_1,n_1|n_0) P(n_1) P(n_0)
\end{align}
where $P(n_0)$ and $P(n_1)$ are the initial distributions of promoter state at time $t_0=-\tau_e$ and $t_0+t_1^{(3)}$ respectively. Assuming promoters at steady-state, both distributions are then given by Eq. (\ref{equ:steady_state_multi}) for a multi-gene system. We then perform the following recursion scheme for $i=\{3,...,k\}$:
\begin{align}
P^{(3)}(\tilde{G}_i,G_i,n_i) = \sum_{n_{i-1}} \sum_{g_i=0}^{g_{\rm max}} P_{t_i^{(3)}}(g_i,n_i|n_{i-1}) P^{(3)}(\underbrace{\tilde{G}_i-(k-i+1)g_i}_{\tilde{G}_{i-1}},\underbrace{G_i-g_i}_{G_{i-1}},n_{i-1})
\end{align}
where $g_{\rm max} = \min{(\lfloor \tilde{G}_i/(k-i+1) \rfloor,G_i)}$.

\paragraph{Step 2: calculate the 5' contribution}

The initial distribution is given by:
\begin{equation}
P^{(5)}(\tilde{G}_1,n_1|n_0) \equiv P^{(5)}(k g_1,n_1|n_0) = P_{t_1^{(5)}}(g_1,n_1|n_0)
\end{equation}
We then perform the following recursion scheme for $i=\{2,...,k\}$:
\begin{equation}
P^{(5)}(\tilde{G}_i,n_i|n_0) = \sum_{n_{i-1}} \sum_{g_i=0}^{g_{\rm max}} P_{t_i^{(5)}}(g_i,n_i|n_{i-1}) P^{(5)}(\underbrace{\tilde{G}_i-(k-i+1)g_i}_{\tilde{G}_{i-1}},n_{i-1}|n_0)
\end{equation}
where $g_{\rm max} = \lfloor \tilde{G}_i/(k-i+1) \rfloor$. Lastly, we sum out $n_k$:
\begin{equation}
P^{(5)}(\tilde{G}_k|n_0) = \sum_{n_k} P^{(5)}(\tilde{G}_k,n_k|n_0)
\end{equation}

\paragraph{Step 3: combine 3' and 5' contributions}

The final joint distribution of 5' and 3' activity in probe space is then given by:
\begin{equation}
P(\tilde{G}^{(5)},\tilde{G}^{(3)}) = \sum_n \sum_{G=0}^{G_{\rm max}} P^{(5)}(\tilde{G}^{(5)}-kG|n) P^{(3)}(\tilde{G}^{(3)},G,n) 
\end{equation}
where $G_{\rm max} = \lfloor \tilde{G}^{(5)}/k \rfloor$. $P^{(3)}$ and $P^{(5)}$ are the joint distributions computed at step 1 and 2. Since the actual signal resolution is of the order of 1 cytoplasmic unit (a fully tagged transcript with $k$ fluorescent probes), the joint distribution can be coarse-grained by aggregating the states $\tilde{G}$ by a block of size $k$ corresponding to a single cytoplasmic unit. The coarse-grained distribution will be denoted $P(G^{(5)},G^{(3)})$ in the following.

\subsection{Likelihood and inference}

\label{sec:inf}

The likelihood of the data $D=\{S^{(5)},S^{(3)}\}$ given the parameters $\theta=(k_{\rm on},k_{\rm off},k_{\rm ini},k_{\rm elo})$ can be expressed in terms of the measurement noise model $P(S^{(5)},S^{(3)}|G^{(5)},G^{(3)})$ (Eq. \ref{equ:noise_measure}) and the joint distribution $P(G^{(5)},G^{(3)}|\theta)$:
\begin{equation}
P(D|\theta)  = \prod_{i=1}^{N_{D}} \sum_{G^{(5)},G^{(3)}} P(S^{(5)}_i,S^{(3)}_i|G^{(5)},G^{(3)}) P(G^{(5)},G^{(3)}|\theta)
\end{equation}
where $N_D$ is the total amount of data. Following a Bayesian approach, the parameters are estimated from the posterior distribution:
\begin{equation}
P(\theta|D) = \frac{P(D|\theta)P(\theta)}{P(D)\equiv \int P(D|\theta)P(\theta) {\rm d}\theta}
\end{equation}
The posterior distribution $P(\theta|D)$ was sampled using a Markov chain Monte Carlo (MCMC) algorithm and the parameters were estimated as the mean of the posterior distribution. We used log-uniform priors as uninformative priors for the rate parameters $(k_{\rm on},k_{\rm off},k_{\rm ini})$ and set the elongation rate $k_{\rm elo}$ to the experimentally measured value of 1.5kb/min \citep{Garcia:2013ha}. At steady-state, a known value of $k_{\rm elo}$ is required to set the temporal scale of the other transcriptional parameters. This can be seen by inspecting of the expressions for the various moments of the nascent transcript distribution (Eq. \ref{equ:2s_mean}, \ref{equ:2s_var}, \ref{equ:k3} and \ref{equ:k4}). Since all moments can be parametrized by the three independent parameters $g_0=k_{\rm ini}/k_{\rm elo}$, $\avg{n}=k_{\rm on}/(k_{\rm on}+k_{\rm off})$ and the ratio $\tau_e/\tau_n=(k_{\rm on}+k_{\rm off})/k_{\rm elo}$, it follows that the model is not identifiable when the temporal scale is not set.

\subsection{Performance}

In order to validate the calculation of the joint distribution and asses the performance of our inference, we first tested the method on synthetic data. Using the Gillespie algorithm \citep{Gillespie:1977dc}, we generated simulated nuclei activity data based on 4 independent gene copies modeled by the telegraph model. We used the probe configuration and gene length of \emph{hb} and assumed a typical elongation rate of 1.5kb/min \citep{Garcia:2013ha}. Measurement noise was included in the simulated data according to the characterization performed previously on real data (cf. Section \ref{sec:error}).

We investigated different parameter regimes and modulation schemes of the mean activity $\mu \equiv \avg{g}$, to test whether the input parameters used to generate the data could be inferred properly (Fig. \ref{fig:S5}A-D). Namely, we tested: 1) modulation of the initiation rate $k_{\rm ini}$ alone with $\tau_n=2$ min and $\avg{n}=0.35$ (cyan dash line), 2) modulation of the on-rate $k_{\rm on}$ alone with $k_{\rm ini}=7$ min$^{-1}$ and $k_{\rm off}=0.25$ min$^{-1}$ (green dash line), 3) modulation of the off-rate $k_{\rm off}$ alone with $k_{\rm ini}=7$ min$^{-1}$ and $k_{\rm on}=0.25$ min$^{-1}$ (blue dash line), 4) modulation of the mean occupancy $\avg{n}$ alone with $k_{\rm ini}=7$ and $\tau_n=2$ min (red dash line). 

For each scenario, we generated 8 batches of data covering the range of normalized activity $\mu/\mu_0$. Each batch was made of 10 independently sampled datasets of 500 nuclei activity measurements. We performed the inference on each dataset individually and reported the mixture of posterior distribution over the 10 datasets to take into account the error due to the randomness of the sampled data. We conclude that the inference framework performs well, since all the inferred quantities cover the true values within error bars. Overall, the inference allows us to distinguish the different tested modulation strategies without ambiguities.

\subsection{Effect of elongation rate on inference}

As we discussed in the previous subsection \ref{sec:inf}, the elongation rate $k_{\rm elo}$ sets the temporal scale of the transcriptional parameters, thus a different elongation rate would lead to different values of the parameters. In the manuscript, we used a value of $k_{\rm elo}=1.5$ kb/min which we previously measured \citep{Garcia:2013ha}. A recent study suggests that this value might be overall larger in the blastoderm embryo, of the order of 2.5 kb/min \citep{Fukaya2017}. We thus sought to determine to which extent this new value would affect our results.

In principle, a different value of $k_{\rm elo}$ rescales the transcriptional parameters in a very predictable way. No matter the elongation rate, the three quantities $g_0$, $\avg{n}$ and $\tau_e/\tau_n$ should be perfectly identifiable. It follows that the new parameters (denoted by the $*$ superscript) have to satisfy the following equations:
\begin{align}
k_{\rm ini} \frac{k_{\rm elo}^{*}}{k_{\rm elo}} &= k_{\rm ini}^{*} \\
\avg{n} &= \avg{n}^{*}  \\
\tau_n \frac{k_{\rm elo}}{k_{\rm elo}^{*}} &= \tau_n^{*}
\end{align}
Inferring the transcriptional parameters from the data with $k_{\rm elo}=2.5$ kb/min instead of $k_{\rm elo}=1.5$ kb/min (as in the main text) confirms the rescaling above (Fig. \ref{fig:S7}). As predicted, $k_{\rm ini}$ and $\tau_n$ are rescaled by a factor $2.5/1.5 \approx 1.67$ and $1.5/2.5=0.6$ respectively, whereas $\avg{n}$ is conserved.


\end{document}